\documentclass{article}
\pdfoutput=1
\usepackage[longnamesfirst]{natbib}
\usepackage{amsmath,amssymb, amsthm, bbm, mathtools} 

\begin{document}

\newcommand{\defeq}{\vcentcolon=}
\newcommand{\eqdef}{=\vcentcolon}

\newcommand{\ind}{\mathbbm{1}}
\newcommand{\forallsp}{ \; \forall \; }
\newcommand{\semisp}{ \; ; \;}

\setlength{\parindent}{0.5cm}

\title{Bayesian inference of Gaussian mixture models with noninformative priors}

\author{Colin J. Stoneking\footnote{Seminar for Statistics, ETH Z\"urich. Email: \texttt{cjstoneking@gmail.com}}}

\maketitle

\begin{abstract}
This paper deals with Bayesian inference of a mixture of Gaussian distributions. A novel formulation of the mixture model is introduced, which includes the prior constraint that each Gaussian component is always assigned a minimal number of data points. This enables noninformative improper priors such as the Jeffreys prior to be placed on the component parameters. We demonstrate difficulties involved in specifying a prior for the standard Gaussian mixture model, and show how the new model can be used to overcome these. MCMC methods are given for efficient sampling from the posterior of this model.
\end{abstract}


\section{Introduction}

Gaussian mixture models (GMMs) are very flexible models with a range of applications, including clustering and approximation of multimodal densities. Bayesian methods are useful for fitting these models to data, because they enable the uncertainty in the model parameters to be directly quantified - by simply examining the posterior distribution or by computing credible intervals. However, it is difficult to make an objective choice of prior for the parameters of the Gaussian components (i.e. their means and variances, in one dimension), when no information is available on which a subjective prior could be based. The typical objective approach would be to use a noninformative prior, i.e. a prior selected according to a formal rule \citep{kass_selection_1996}. For GMMs, this approach is usually not possible. Standard noninformative priors such as the Jeffreys prior \citep{jeffreys_theory_1961, kass_selection_1996} generally cannot be used for mixture models, because they tend to be improper, and placing independent improper priors on the parameters of each mixture component will cause the posterior to be improper as well \citep{roeder_practical_1997, stephens_bayesian_1997, marin_bayesian_2005}.\\
\indent Given this difficulty, one popular approach for GMMs has been to use proper priors, with their parameters chosen so that they are ``weakly informative'' \citep{richardson_bayesian_1997}. Heuristically, this can be defined as follows: the prior densities should be relatively flat in the range of values that the parameters could be expected to take, given the range of the data \citep{raftery_hypothesis_1996}. Such priors are also referred to as ``locally uniform'' \citep{box_bayesian_1973} or as ``diffuse'' \citep{kass_selection_1996}. For example, weakly informative priors were used in \citep{ferguson_bayesian_1983, raftery_hypothesis_1996, richardson_bayesian_1997, stephens_bayesian_2000}.\\
\indent In some cases, weakly informative priors can be justified as an objective approach by the fact that as they are made increasingly weak, the posterior density converges to the density that is obtained with some noninformative improper prior. For example, in hierarchical models, this is the case for uniform$(0, L)$ prior densities on the standard deviations of group level effects: as $L \to \infty$, under some conditions, one obtains the same posterior as if an improper uniform$(0, \infty)$ prior had been used \citep{gelman_prior_2006}. However, this convergence cannot occur for mixture models, as the posterior is improper if the priors are improper. In other settings where there is no proper limiting posterior, weakly informative priors are prone to issues such as sensitivity of the posterior to prior parameters, and can give nonsensical posteriors \citep{kass_selection_1996, berger_bayesian_2000}. Therefore, we might expect weakly informative priors to lead to practical problems in GMMs as well. In this paper, we show that this is indeed the case, and propose a straightforward modification of the mixture model which solves this problem.\\ 
\indent The paper is organized as follows: in section \ref{section_standard}, after introducing the standard approach for Bayesian inference with a GMM, we show that for a simple example data set, weakly informative priors are prone to a severe prior domination effect. Because of this, there is no generally valid way to choose the prior parameters when attempting to use weakly informative priors. In section \ref{section_min}, we show that a slight modification of the standard GMM allows noninformative priors to be used. This avoids the problem of parameter choice. In section \ref{Section_imp}, we provide MCMC implementations of our model and compare it with the standard model on real and simulated data.

\subsection{Related work}

Various approaches have been proposed for placing priors on the component parameters of a GMM. These can be roughly divided into three strategies. The first approach is to use proper priors, with the prior parameters chosen such that the prior is suitably weakly informative. One disadvantage of this approach is the fact that multiple prior parameters usually need to be specified. For example, the model of \cite{richardson_bayesian_1997} has 4 parameters related to scale or shape, for which no default values are available. \citet{richardson_bayesian_1997} propose heuristic values for these based on the range of the data values. In this paper, we demonstrate a further, serious problem with weakly informative priors (see section \ref{section_weakinf}).\\
\indent An alternative is to use ``partially proper'' priors which are noninformative in some specific way, similar to the improper priors which would be available in a non-mixture setting \citep{mengersen_testing_1996, roeder_practical_1997}. These priors have been shown to give proper posteriors. However, they still require some rather crucial information to be specified. For example, \citet{mengersen_testing_1996} developed a prior in which the means of the mixture components are specified in terms of their differences from each other. The prior on the overall location of the mixture density can then be improper. However, a proper prior has to be used for the differences of the component means. This is an important feature of the model, and the fact that one must base it on subjective input is problematic. The prior proposed by \cite{roeder_practical_1997} follows a similar approach.\\
\indent Finally, one can use an improper prior, and employ a modified sampling algorithm that makes the posterior proper, by forcing each component to always have a minimal number of data points assigned to it \citep{diebolt_estimation_1994}. This has been shown to be equivalent to multiplying the original priors with a data-dependent factor \citep{wasserman_asymptotic_2000}.  
The advantage of this approach is that there are no subjective choices to make. The disadvantage is that the prior becomes data-dependent, which is formally incorrect in a Bayesian framework. \cite{wasserman_asymptotic_2000} motivated the data-dependent prior primarily by showing that it leads to intervals with second-order correct frequentist coverage.\\
\indent Our approach is related to the work of \cite{diebolt_estimation_1994} and \cite{wasserman_asymptotic_2000}, in that our method enables improper priors to be used by ensuring that each component is assigned a minimum number of data points. However, our modification does not result in any data-dependence of the priors. Instead, our approach is to recast inference in terms of a slightly modified model.

\section{Bayesian inference for GMMs}\label{section_standard}

In this section, we introduce the GMM as it is typically used in Bayesian inference. For simplicity, we focus on the 1-dimensional case (the generalization to more than one dimension is straightforward). Thus, suppose we have a sample $\mathbf{x} = (x_1, \dots x_N) \in \mathbb{R}^N$ of $N$ data points, each in $\mathbb{R}$. We assume throughout this paper that these data points are i.i.d. samples from some (unknown) distribution that is dominated by Lebesgue measure on $\mathbb{R}$. We want to model their density with a mixture of $K$ univariate Gaussian densities, where $K$ is specified in advance, and fixed. According to this model, the $x_j, j \in \{1, \dots N\}$ are i.i.d., each following the mixture density given by:
\begin{equation}\label{mixture_1D}
f(x | \boldsymbol{\mu}, \boldsymbol{\sigma}, \mathbf{p}) = \sum_{i=1}^K p_i f_{\mathcal{N}}(x\; ; \mu_i, \sigma^2_i)\\
\end{equation}
where $f_{\mathcal{N}}(\cdot\; ; \mu, \sigma^2)$ denotes the univariate Gaussian density function with mean $\mu$ and variance $\sigma^2$. The parameters of the component densities thus consist of $\boldsymbol{\mu} = (\mu_1, \dots, \mu_K) \in \mathbb{R}^K$ and $\boldsymbol{\sigma}^2 = (\sigma_1^2, \dots, \sigma_K^2) \in \mathbb{R}_{+}^K$. The mixture weights $\mathbf{p} = (p_1, \dots p_K)$ must satisfy:
\begin{align*}
& p_i >  0 \;\forall\; i \in 1 \dots K\\
& \sum_{i=1}^K p_i = 1
\end{align*}
Inference for this model is greatly simplified by putting it in a generative representation, with the aid of latent variables that indicate which component generated which data point. Let $\boldsymbol{\mathcal{G}} \defeq \{1, \dots K\}^N$ be the set of all possible assignments of the $N$ data points to $K$ components. Note that we can rewrite the likelihood from model \eqref{mixture_1D} as:
\begin{align*}
f(\mathbf{x}|\boldsymbol{\mu}, \boldsymbol{\sigma}, \mathbf{p} ) = & \prod_{j=1}^N \sum_{i=1}^K p_i f_{\mathcal{N}}(x_j\; ; \mu_i, \sigma^2_i)\\
 = & \sum_{\mathbf{G} \in \boldsymbol{\mathcal{G}}} \prod_{i=1}^K \prod_{j: G_j = i} p_i f_{\mathcal{N}}(x_j\; ; \mu_i, \sigma^2_i)\\
  = & \sum_{\mathbf{G} \in \boldsymbol{\mathcal{G}}} f(\mathbf{x} | \boldsymbol{\mu}, \boldsymbol{\sigma}, \mathbf{G}) f(\mathbf{G} | \mathbf{p})\\
\end{align*}
where $f(\mathbf{x} | \boldsymbol{\mu}, \boldsymbol{\sigma}, \mathbf{G}) = \prod_{i=1}^K \prod_{j: G_j = i} f_{\mathcal{N}}(x_j\; ; \mu_i, \sigma^2_i)$ and $f(\mathbf{G} | \mathbf{p}) = \prod_{i=1}^K \prod_{j: G_j = i} p_j$. 
This means that \eqref{mixture_1D} is equivalent to a two-stage generative model where, to generate a value of $x_j$, we first draw a value of a latent variable $G_j$ distributed on $\{1,\dots K\}$ with probabilities $\mathbf{p}$, and then draw a value of $x_j$ from $f_{\mathcal{N}}(\cdot\; ; \mu_{G_j}, \sigma_{G_j}^2)$. For the purposes of inference, we now assume that $\mathbf{x}$ was generated by such a model, and take $\mathbf{G} = (G_1, \dots G_N)$ to be the vector of latent variables associated with $\mathbf{x}$.\\
\indent In a Bayesian framework, $\mathbf{p}$, $\boldsymbol{\mu}$, $\boldsymbol{\sigma}^2$, $\mathbf{G}$ and $\mathbf{x}$ are all treated as random variables. The prior distribution of $\mathbf{p}$ is generally taken to be the Dirichlet distribution of order $K$ with parameters $\boldsymbol{\delta} = (\delta_1,\dots,\delta_K)$ \citep{diebolt_estimation_1994, wasserman_asymptotic_2000}. Often, $\boldsymbol{\delta} = (1, \dots, 1)$ is chosen, which gives a prior on $\mathbf{p}$ that is uniform over the probability simplex. This choice will be made throughout this paper. We refer to the model given by \eqref{mixture_1D} with a Dirichlet prior on $\mathbf{p}$ as the standard GMM, reflecting the fact that many models found in the literature include this basic structure (as long as $\mathbf{p}$ is random, and not fixed a priori).\\


\subsection{Weakly informative priors}\label{section_weakinf}

\begin{figure}
\vspace*{-3cm}\hspace*{-2cm}\includegraphics[scale=0.75,clip, trim = 0mm 110mm 0mm 100mm]{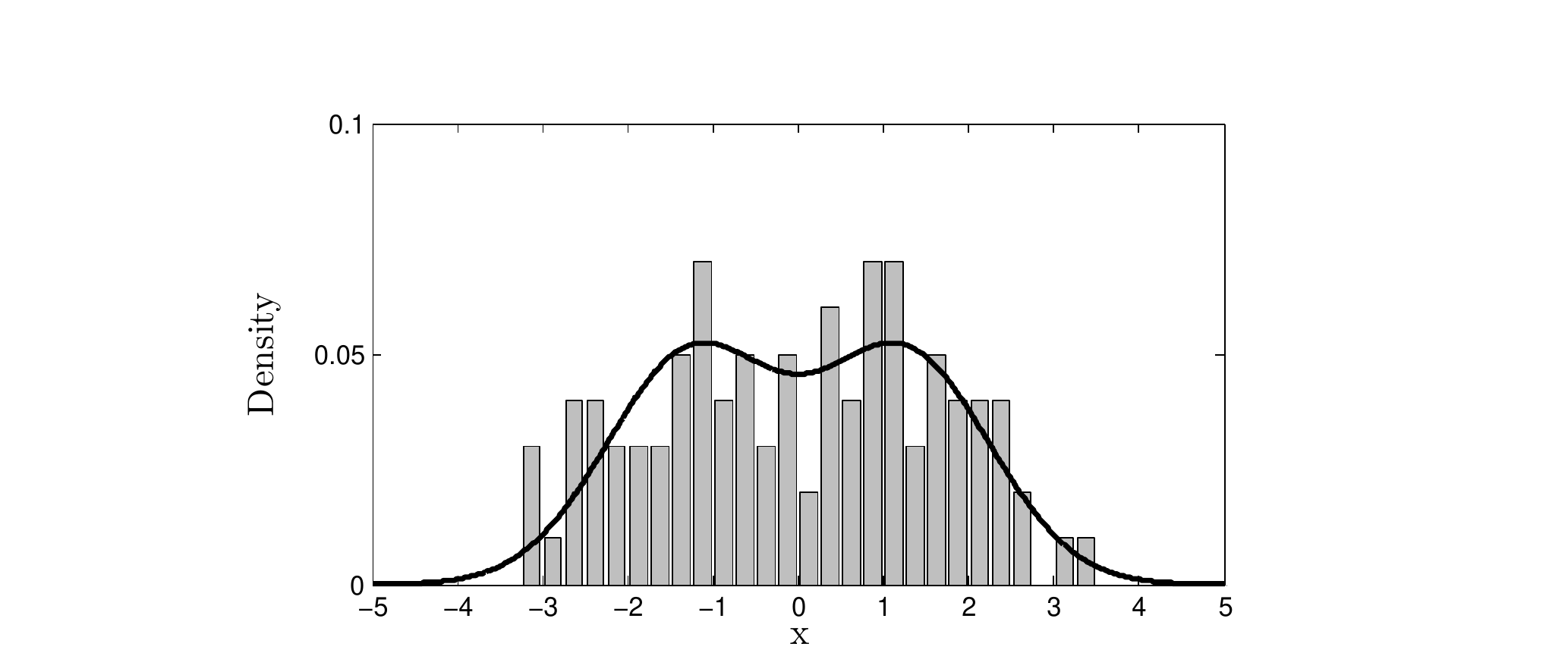}
\caption{Normalized histogram of the sample from the mixture density \eqref{synth_density}, with scaled density superimposed.}\label{synth_data}
\end{figure}

\begin{figure}
\vspace*{-2cm}\hspace*{-2cm}\includegraphics[scale=0.75,clip, trim = 0mm 90mm 0mm 70mm]{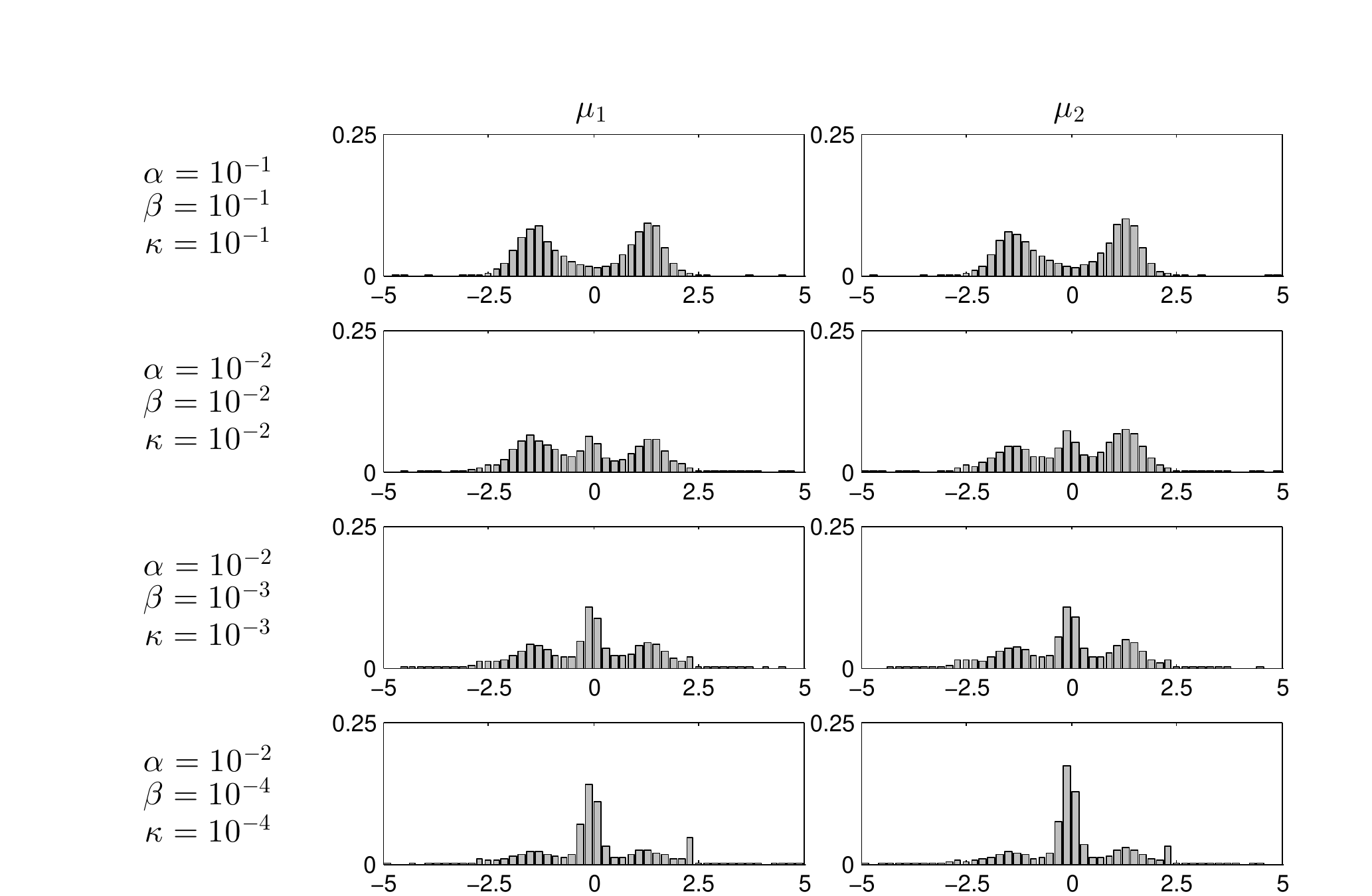}
\caption{Samples from the posterior of $\mu_1$ and $\mu_2$, for the data shown in figure \ref{synth_data}, using the GMM \eqref{mixture_1D} with a normal-inverse gamma prior, with parameters given on the left. The Markov chain was simulated for an initial period of $10^4$ steps, which were discarded (burn-in), then a further $10^5$ steps were run, and every 10-th of these was saved. $\alpha$ was never made smaller than 0.01, because otherwise numerical problems occur if a component is assigned no data points, and we must sample its variance from the inverse gamma prior.}\label{synth_data_nivg}
\end{figure}

\begin{figure}
\vspace*{0cm}\hspace*{-5cm}\includegraphics[angle=90,scale=0.75,clip, trim = 95mm 0mm 90mm 0mm]{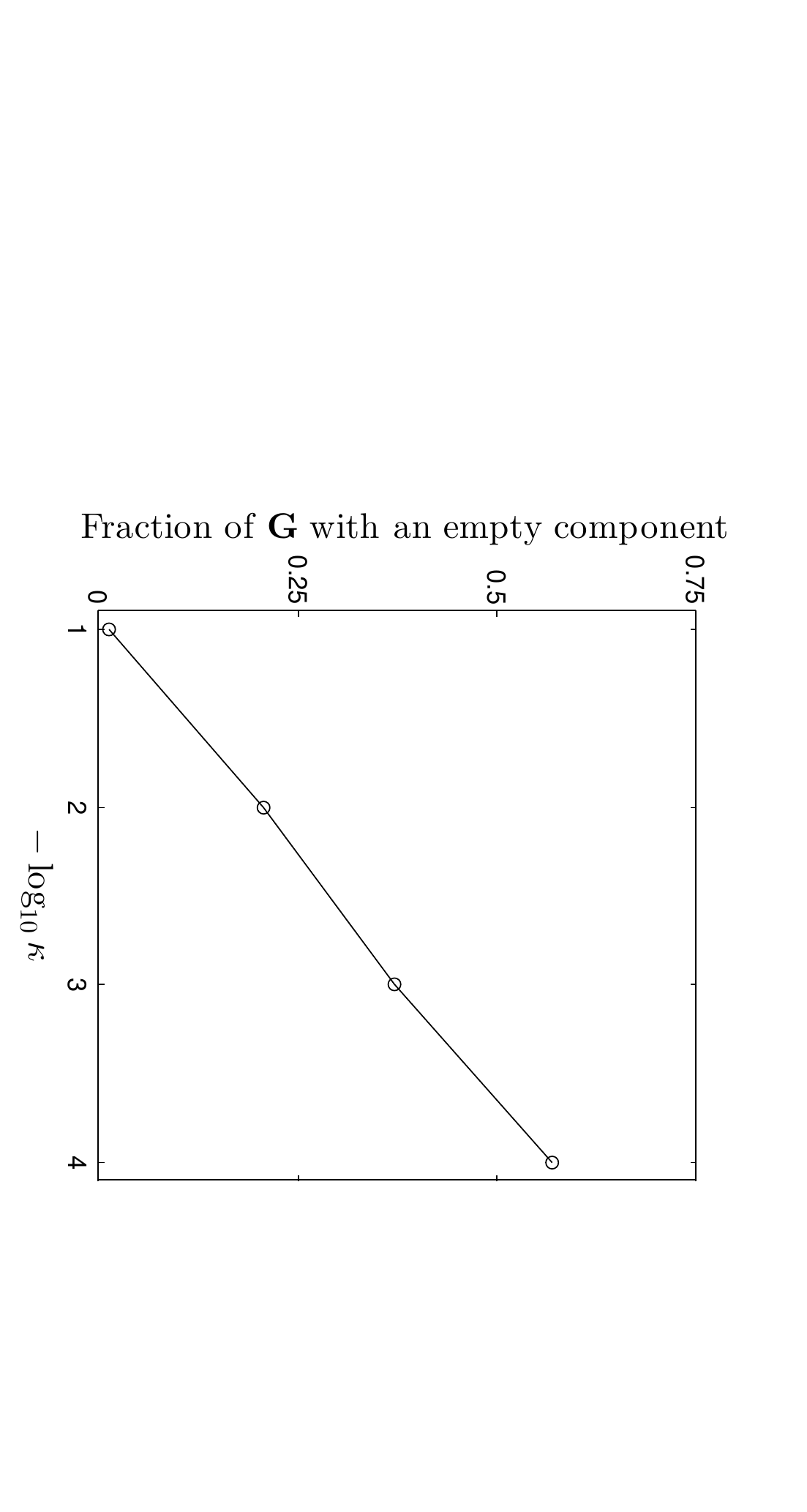}
\caption{Fraction of sampled $\mathbf{G}$ which gave one component no data points, from the inference shown in figure \ref{synth_data_nivg}.}\label{synth_data_nivg_zeros}
\end{figure}

Different options are available for the prior distribution $\pi(\boldsymbol{\mu}, \boldsymbol{\sigma}^2)$. The usual choice is to assume prior independence between the component densities, and then to place a proper prior $\pi(\mu_i, \sigma_i^2)$ on each pair $(\mu_i, \sigma_i^2), i \in \{1, \dots K\}$. When proper priors are used, their parameters are often chosen such that the priors are weakly informative, i.e.\ with relatively flat densities over the range of relevant values. However, in practice, weakly informative priors can strongly constrain the posterior such that the result of the inference becomes very poor. As an example, consider a model with a conjugate normal-inverse gamma prior:
\begin{align*}
& \pi(\mu_i, \sigma_i^2) = f_{\mathcal{N}}\left(\mu_i\semisp 0, \frac{\sigma_i^2}{\kappa} \right) \frac{\beta^\alpha}{\Gamma(\alpha)} \left(\sigma_i^{-2}\right)^{\alpha + 1} \exp\left(- \sigma_i^{-2}\beta\right)\\
& i \in \{1, \dots K\}\\
& \alpha > 0,\; \beta > 0,\; \kappa > 0
\end{align*}
Here the Gaussian part of the density has a zero mean for simplicity; it is of course also possible for it to be non-zero. We assume that the data are approximately centered, so that the zero mean makes sense. Conventionally, this prior can be made weakly informative by setting the parameters $\alpha$, $\beta$ and $\kappa$ to take similar, low values (e.g. they could all be set to 0.01). The idea is that the inverse-gamma part approximates the Jeffreys prior for $\sigma^2_i$ \citep{lunn_bugs_2012} (which means it is relatively flat on the log scale), and the conditional Gaussian part is approximately flat because of the small value of $\kappa$. Obviously this is a heuristic approach, and the precise values will depend on the scale of the data.\\
\indent To demonstrate a problem inherent to such priors, we use a synthetic dataset, consisting of 100 datapoints sampled from the following two-component GMM:
\begin{equation}\label{synth_density}
f(x) = \frac{1}{2}f_{\mathcal{N}}\left(x \; ;\; -1.25, 1\right) + \frac{1}{2}f_{\mathcal{N}}\left(x \; ;\; 1.25, 1\right)
\end{equation}
The two component densities overlap strongly. Figure \ref{synth_data} shows a histogram of the data, with the mixture density superimposed.
%
%
Figure \ref{synth_data_nivg} shows histograms of approximate samples from the posterior distributions of $\mu_1$ and $\mu_2$, given these data, as the values of $\alpha$, $\beta$ and $\kappa$ are decreased. These were computed by Gibbs sampling, using a standard scheme which alternated between sampling new values for $\mathbf{G}$ conditional on $(\boldsymbol{\mu}, \boldsymbol{\sigma}^2, \mathbf{p})$, and new values for $(\boldsymbol{\mu}, \boldsymbol{\sigma}^2, \mathbf{p})$ conditional on $\mathbf{G}$. This particular form of Gibbs sampling is also referred to as data augmentation \citep{tanner_calculation_1987, diebolt_estimation_1994}.\\
\indent Posterior densities for the component parameters of mixture densities generally can have multiple modes \citep{celeux_computational_2000, marin_bayesian_2005}. This is also evident in figure \ref{synth_data_nivg} - the histograms of $\mu_1$ and $\mu_2$ in the top row are markedly bimodal. These modes arise because the posterior is invariant under permutations of the component indices. As a result, the samples of $\mu_1$ effectively contain contributions from both mixture components, and similarly for $\mu_2$, a phenomenon referred to as label-switching. For further inference, a variety of methods would be available to separate the samples from the two ``true'' components \citep{stephens_dealing_2000, hurn_estimating_2003, jasra_markov_2005, grun_dealing_2009, yao_bayesian_2009}. For our purposes, the posteriors are sufficient as they are. Convergence of the sampler was assessed by the fact that the histograms for $\mu_1$ and $\mu_2$ are very similar - this indicates that the sampler was able to move well between the different symmetric modes of the posterior \citep{celeux_computational_2000, lee_bayesian_2008}.\\
\indent When $\alpha$, $\beta$ and $\kappa$ are all equal to 0.1, the posterior distributions of $\mu_1$ and $\mu_2$ each have two modes, which reflect the fact that the data can be more or less well separated into a group with a mean of approximately $-1.25$, and a group with a mean of $1.25$. As the prior parameters are decreased, a central mode appears, and eventually dominates the posterior. This mode is produced by assignments of the data points to components such that one component takes all points or a large majority, and thus has a posterior mean of approximately zero. We can see this by plotting the proportion of samples of $\mathbf{G}$ in which one of the components is assigned no points - this proportion increases steadily as the prior parameters $\alpha$, $\beta$ and $\kappa$ decrease (figure \ref{synth_data_nivg_zeros}). Therefore, varying the prior so it is supposedly less informative actually constrains the posterior so that the central mode plays a larger and larger role. This is a prior domination effect \citep{kass_selection_1996}, because the prior parameters, not the data, control the contribution of the central mode to the posterior.\\
\indent This effect can be explained using the explicit solution for the posterior probability of the latent variables $\mathbf{G}$. From Bayes' formula, it is given by:
\begin{equation}\label{z_post}
f(\mathbf{G}|\mathbf{x}) \propto \left(\prod_{i=1}^K  \int \int \pi(\mu_i, \sigma^2_i) \prod_{j:G_j = i} f_\mathcal{N}(x_j \semisp \mu_i, \sigma^2_i)\; \mathrm{d} \mu_i \mathrm{d} \sigma^2_i \right) \pi(\mathbf{G})
\end{equation}
In this specific case, $\pi(\mu_i, \sigma^2_i)$ is the normal-inverse gamma prior, and $\pi(\mathbf{G})$ is the prior on $\mathbf{G}$ given the Dirichlet prior $\pi_D(\mathbf{p})$, which can be obtained by integrating out $\mathbf{p}$, and is given by:
\begin{equation}\label{standard_z_prior}
\pi(\mathbf{G}) = \int f(\mathbf{G}|\mathbf{p})\pi_D(\mathbf{p})\mathrm{d}\mathbf{p} = \frac{\Gamma\left(\sum_{i=1}^K \delta_i\right)}{\Gamma\left(N + \sum_{i=1}^K\delta_i\right)}\prod_{i=1}^K \frac{\Gamma\left(n_i(\mathbf{G}) + \delta_i\right)}{\Gamma(\delta_i)}
\end{equation}
where $n_i(\mathbf{G})$ is the number of points assigned to the $i$-th component, i.e. 
\begin{equation}\label{n_i}
n_i(\mathbf{G}) = \#\{G_j : G_j = i\},\;\; i \in \{1, \dots K\}
\end{equation}\\
We use the following short notation for the integrals:
\begin{equation}\label{fi_notation}
f_i(\mathbf{x},\mathbf{G}) \defeq \; \int \int \pi(\mu_i, \sigma^2_i)\prod_{j:G_j = i}  f_\mathcal{N}(x_j \semisp \mu_i, \sigma^2_i) \; \mathrm{d} \mu_i \mathrm{d} \sigma^2_i\\
\end{equation}
so that
\begin{align*}
f(\mathbf{G}|\mathbf{x}) \propto \pi(\mathbf{G}) \prod_{i=1}^K f_i(\mathbf{x} , \mathbf{G}) 
\end{align*}
A closed-form expression for $f_i(\mathbf{x},\mathbf{G})$ can be obtained by straightforward integration:
\begin{equation}\label{fi_expression}
f_i(\mathbf{x},\mathbf{G})= \; \frac{(2\beta)^\alpha \kappa^{\frac{1}{2}}\;\Gamma\left(\frac{n_i}{2} + \alpha\right)}{\pi^{\frac{n_i}{2}} (n_i + \kappa)^{\frac{n_i + 1}{2} + \alpha}\;\Gamma(\alpha)} \left( \frac{1}{n_i + \kappa}\sum_{j:G_j=i} x_j^2 -\left(\frac{1}{n_i + \kappa}\sum_{j:G_j=i} x_j\right)^2 + \frac{2\beta}{n_i + \kappa}\right)^{-\frac{n_i}{2} -\alpha}\\
\end{equation}
Note that this is simply equal to 1 for $n_i=0$. Now assume for simplicity that we use a parametrization in which the prior parameters are related by fixed linear functions. We then can show that as the prior becomes less informative, the posterior density $f(\mathbf{G}|\mathbf{x})$ will become concentrated on those $\mathbf{G} \in \boldsymbol{\mathcal{G}}$ which assign all the data points to one component:\\

\noindent \textbf{Lemma 1}\\

\noindent Assume that $N > K$, and that $\alpha = c_1 \kappa$ and $\beta = c_2 \kappa$, with fixed constants $c_1 > 0$ and $c_2 > 0$. Let $\mathbf{G'}, \mathbf{G''} \in \boldsymbol{\mathcal{G}}$ be vectors of latent variables. If $\mathbf{G'}$ assigns all data points to a single component, i.e. $\#\{i \in \{1, \dots K\} : n_i(\mathbf{G'}) = 0\} = K - 1$, and $\mathbf{G''}$ does not do this, then $P-a.s.$:

\begin{displaymath}
\lim_{\kappa \to 0} \frac{f(\mathbf{G'}|\mathbf{x})}{f(\mathbf{G''}|\mathbf{x})} = \infty
\end{displaymath}\\

\noindent The proof is given in the appendix \eqref{lemma_1}. Here, the $P-a.s.$ refers to the (unknown) law on $\mathbf{x}$.
Similarly, one can show that for the precise situation we had in the later part of the simulations, namely $\alpha$ held fixed and $\beta = c_2\kappa$ while $\kappa \to 0$, then if $\mathbf{G'}$ assigns all data points to a single component and $\mathbf{G''}$ gives at least two components each more than one data point, $\lim_{\kappa \to 0}\frac{f(\mathbf{G'}|\mathbf{x})}{f(\mathbf{G''}|\mathbf{x})} = \infty$ also holds.\\
\indent In the case of our example analysis with a mixture of 2 components, this means that as the prior is made less and less informative, eventually most samples from the posterior of $\mu_1$ and $\mu_2$ will be conditional on instances of $\mathbf{G}$ such that one component has all or all but one of the data points assigned to it. The posterior density of $\mu$ for this ``greedy'' component will be centered at the overall mean of the data, approximately $0$. This explains the increase in the mode centered at $0$ which can be seen in figure \ref{synth_data_nivg}.  Figure \ref{synth_data_nivg_zeros} shows the proportion of sampled $\mathbf{G}$ which assign one component no data points at all - this proportion increases as the prior becomes less informative.

\begin{figure}
\vspace*{0cm}\hspace*{-2cm}\includegraphics[scale=0.75,clip, trim = 0mm 90mm 0mm 60mm]{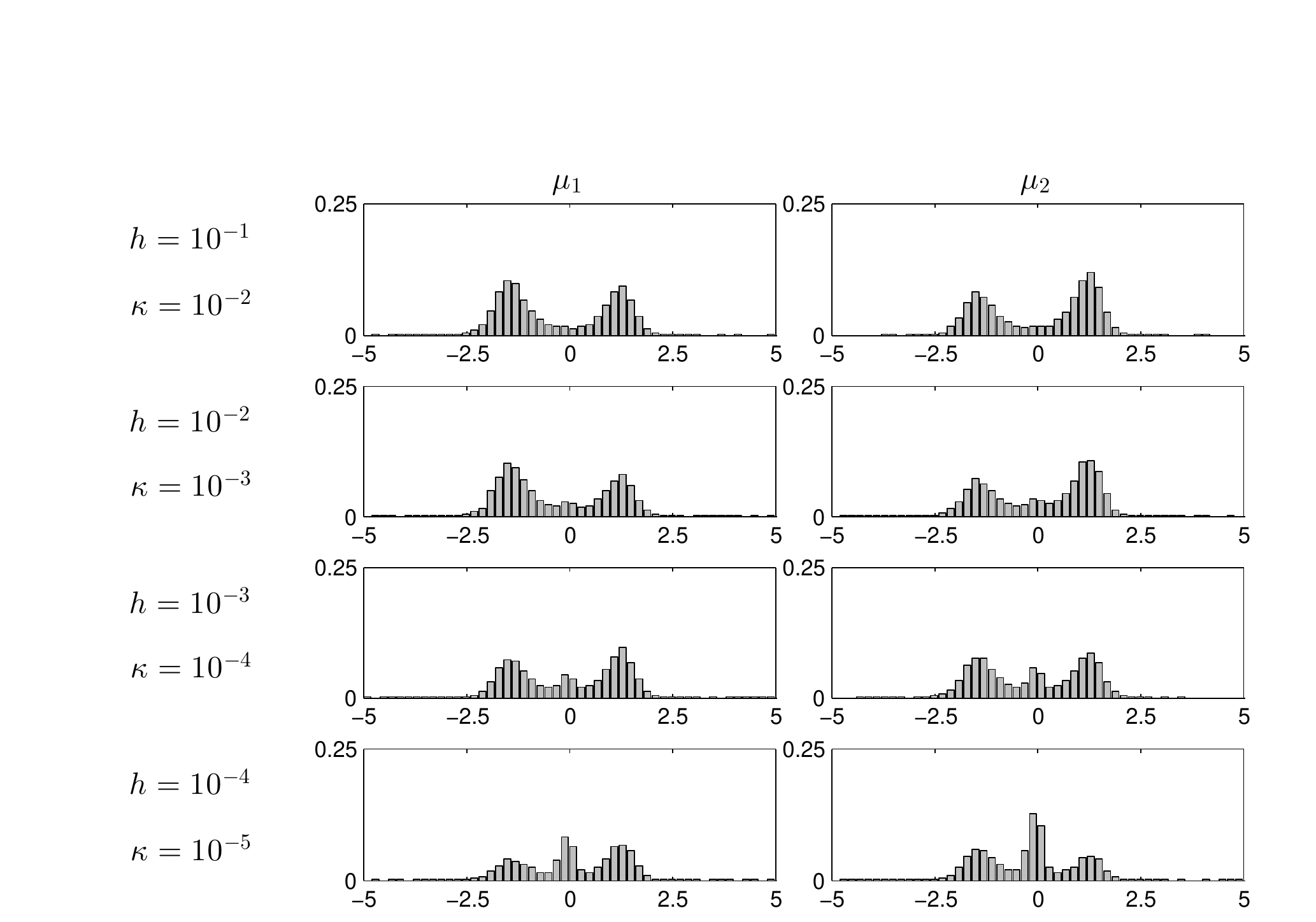}
\caption{Samples from the posterior of $\mu_1$ and $\mu_2$, for the data shown in figure \ref{synth_data}, using the GMM \eqref{mixture_1D} with the priors of \citet{richardson_bayesian_1997}. Burn-in was $10^4$, post-burn-in $10^5$, every 10-th state of the chain was sampled.}\label{synth_data_rg}
\end{figure}

\begin{figure}
\vspace*{-1cm}\hspace*{-5cm}\includegraphics[angle=90,scale=0.75,clip, trim = 95mm 0mm 90mm 0mm]{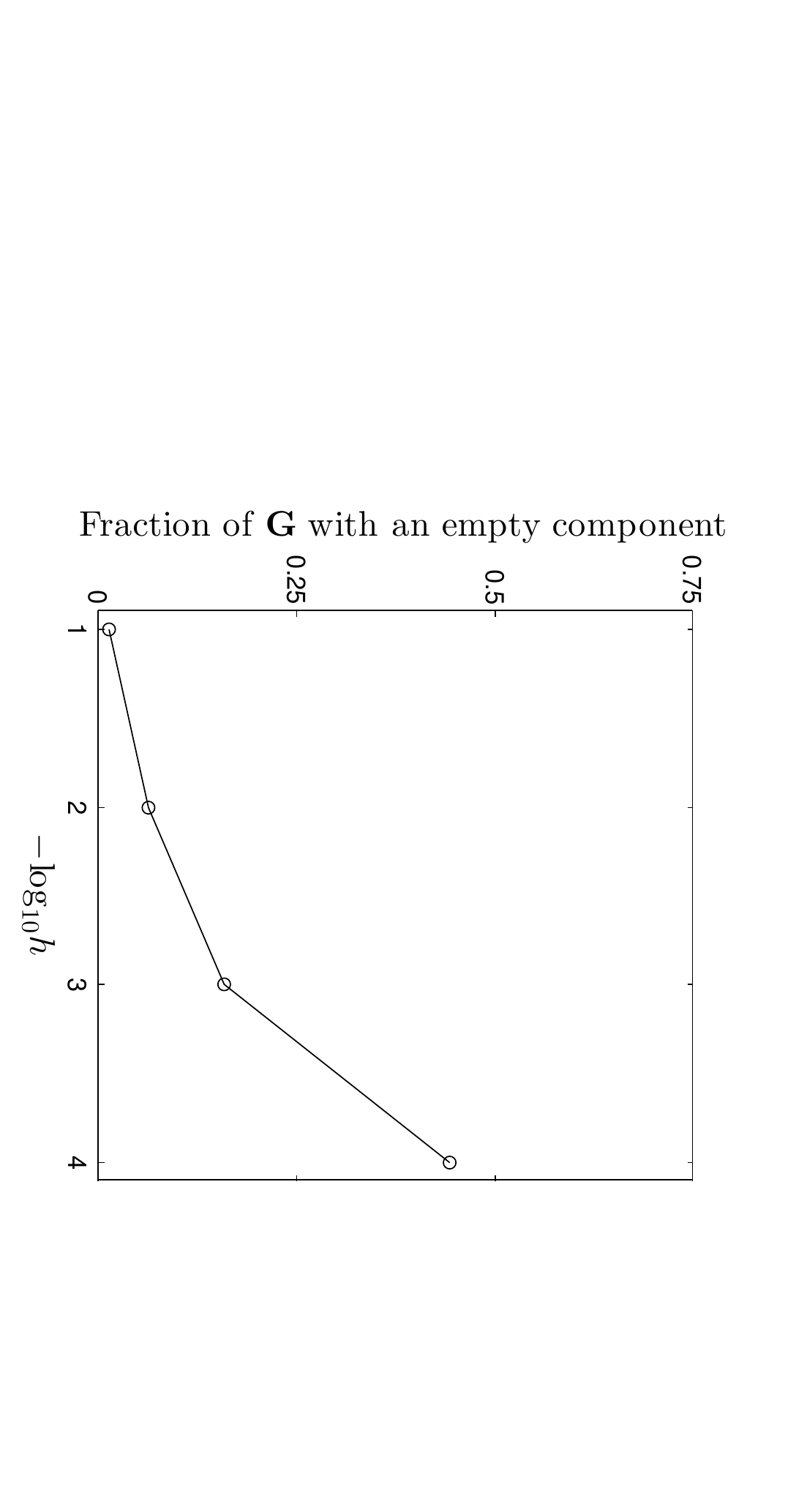}
\caption{Fraction of sampled $\mathbf{G}$ which gave one component no data points, from the inference shown in figure \ref{synth_data_rg}.}\label{synth_data_rg_zeros}
\end{figure}

We chose the model with a conjugate normal-inverse gamma prior as an example because it is possible to find an explicit solution for the posterior $f(\mathbf{G}|\mathbf{x})$. The prior domination effect can also be observed in a model proposed by \citet{richardson_bayesian_1997}. This model is slightly more complicated, as it introduces an additional hierarchical level via the variable $\beta$:
\begin{align*}
&\pi(\mu_i) = f_{\mathcal{N}}\left(\mu_i \semisp 0, \kappa^{-1}\right)\\
&\pi(\sigma_i^2|\beta) =  \frac{\beta^\alpha}{\Gamma(\alpha)} \left(\sigma_i^{-2}\right)^{\alpha + 1} \exp\left(-\beta\sigma_i^{-2}\right)\\
&\pi(\beta) =  \frac{h^g}{\Gamma(g)} \beta^{g-1}\exp\left(-h\beta\right)\\
&i \in \{1, \dots K\}\\ &\alpha > 0,\; g > 0,\; h > 0,\; \kappa > 0
\end{align*}
The original formulation of the model allows for a non-zero mean of $\pi(\mu_i)$; as before, we take it to be zero. We implemented the GMM with these priors using Gibbs sampling. For the parameters, we used $\alpha = 2$, $g = 0.2$ as proposed in \cite{richardson_bayesian_1997}. We also kept $h = 10\kappa$ at all times, to match their settings. We then decreased $h$ and $\kappa$, following the approach in \cite{richardson_bayesian_1997} that these parameters should be small for the prior to be weakly informative.  The result (figures \ref{synth_data_rg}, \ref{synth_data_rg_zeros}) is similar as before. The central mode does not become quite as pronounced, which seems to be because the additional hierarchical level in this prior makes it more robust to parameter variation \citep{robert_bayesian_2007}. However, the mode still changes strongly in size as the prior parameters are varied.

\indent For the model with these priors, proving the equivalent of lemma 1 is more difficult.
We can make the general (informal) argument that for weakly informative priors, the integral $\int \int \pi(\mu_i, \sigma^2_i) \prod_{j:G_j=i}  f_\mathcal{N}(x_j \semisp \mu_i, \sigma^2_i) \mathrm{d} \mu_i \mathrm{d} \sigma^2_i$ will tend to be small for $n_i(\mathbf{G}) > 0$, because the prior $\pi$ will be small for any pair $(\mu_i, \sigma^2_i)$ for which the likelihood is reasonably large. A similar argument was made by \citet{jennison_discussion_1997} in the context of mixture models with a random number of components. Therefore,  we expect that as the prior is made increasingly weak, $f(\mathbf{G}|\mathbf{x})$ will become concentrated on $\mathbf{G}$ that allocate no data points to at least one component. We thus expect to see effects like those in figures \ref{synth_data_nivg} and \ref{synth_data_rg} for all weakly informative priors. \\
\indent This means that there is no well-founded, general method to choose the parameters of a given prior so as to make it weakly informative for a GMM. We cannot simply choose the parameters such that the prior density is extremely diffuse, because this may affect the posterior for $\mathbf{G}$ to such an extent that the posterior for the component parameters is noticeably affected. Crucially, we cannot assume that the presence of a central mode as in figures \ref{synth_data_nivg} and \ref{synth_data_rg} must always be an artefact due to the prior. So we would have to somehow choose the parameters to be at some middle ground between having the priors be sufficiently diffuse and avoiding the prior domination effect on $\mathbf{G}$, but it is not clear how this could be objectively achieved, in general. Note that the example dataset is not particularly special; its only important aspect is that its two modes are fairly close together, which is a feature we could expect many real datasets to have.\\

\section{The GMM with noninformative priors}\label{section_min}

Since weakly informative priors are problematic in practice, we developed an approach that enables noninformative improper priors to be used. The main motivation for this is that noninformative priors do not require parameters to be specified, so they avoid the problems seen in the previous section. It is helpful to first write the posterior $f(\boldsymbol{\mu}, \boldsymbol{\sigma}, \mathbf{p} | \mathbf{x})$ as:
\begin{equation}\label{post_G_expansion}
f(\boldsymbol{\mu}, \boldsymbol{\sigma}, \mathbf{p} | \mathbf{x}) \propto \sum_{\mathbf{G} \in \boldsymbol{\mathcal{G}}}  f(\mathbf{x} | \boldsymbol{\mu}, \boldsymbol{\sigma}, \mathbf{G})f(\mathbf{G} | \mathbf{p}) \pi(\boldsymbol{\mu}, \boldsymbol{\sigma}, \mathbf{p})
\end{equation}
Suppose that $\pi(\boldsymbol{\mu}, \boldsymbol{\sigma}, \mathbf{p})$ is a product of independent Jeffreys priors on the pairs $(\mu_i, \sigma_i)$ and a Dirichlet prior $\pi_D(\mathbf{p})$:
\begin{equation}\label{improper_prior}
\pi(\boldsymbol{\mu}, \boldsymbol{\sigma}, \mathbf{p}) = \pi_J(\boldsymbol{\mu}, \boldsymbol{\sigma})\pi_D(\mathbf{p})
\end{equation}
\begin{equation}\label{Jeffreys_prior}
\pi_J(\boldsymbol{\mu}, \boldsymbol{\sigma}) \propto \prod_{i=1}^K \sigma_i^{-1}
\end{equation}
Then we obtain:
\begin{equation*}\label{orig_posterior}
f(\boldsymbol{\mu}, \boldsymbol{\sigma}, \mathbf{p} | \mathbf{x}) \propto  \sum_{\mathbf{G} \in \boldsymbol{\mathcal{G}}}\left( \prod_{i=1}^K \sigma_i^{-1} \prod_{j:G_j = i}  f_{\mathcal{N}}(x_j \semisp \mu_i, \sigma_i^2) \right) f(\mathbf{G} | \mathbf{p}) \pi_D(\mathbf{p})
\end{equation*}
This posterior is improper, as the terms $\sigma_i^{-1}\prod_{j:G_j = i} f_{\mathcal{N}}(x_j \semisp \mu_i, \sigma_i^2)$ are not integrable over $\{(\mu_i, \sigma_i) \in \mathbb{R} \times \mathbb{R}_{+} \}$ when $n_i(\mathbf{G}) < 2$. However, if $n_i(\mathbf{G}) \geq 2$, then these terms are $P-a.s.$ integrable, where $P$ is the (unknown) law of $\mathbf{x}$ (see appendix, \ref{int_1D}). Therefore, if we exclude from the posterior all $\mathbf{G}$ with $n_i(\mathbf{G}) < 2$ for any $i \in \{1, \dots K\}$, we can use the improper Jeffreys prior and still have a $P-a.s.$ proper posterior. This modified posterior is given by:

\begin{equation}\label{good_posterior}
f^{\star}(\boldsymbol{\mu}, \boldsymbol{\sigma}, \mathbf{p} | \mathbf{x}) \propto \sum_{\mathbf{G} \in \boldsymbol{\mathcal{G}}^{\star}}  f(\mathbf{x} | \boldsymbol{\mu}, \boldsymbol{\sigma}, \mathbf{G}) f(\mathbf{G}|\mathbf{p}) \pi_J(\boldsymbol{\mu}, \boldsymbol{\sigma}) \pi_D(\mathbf{p})
\end{equation}
where $\boldsymbol{\mathcal{G}}^\star$ contains all ``good'' $\mathbf{G}$:
\begin{equation*}
\boldsymbol{\mathcal{G}}^\star = \{ \mathbf{G} \in \boldsymbol{\mathcal{G}} : n_i(\mathbf{G}) \geq 2 \;\forall\; i \in \{1, \dots K\}\}
\end{equation*}
The motivation for this posterior is that it is minimally modified compared to the original \eqref{post_G_expansion} - only some $\mathbf{G}$ are dropped to ensure propriety. This modification was first introduced by \cite{diebolt_estimation_1994}, who applied it but subsequently inferred the model parameters as if the posterior had not been modified. \cite{wasserman_asymptotic_2000} treated it more formally, introducing a data-dependent modification of the prior $\pi(\boldsymbol{\mu}, \boldsymbol{\sigma}, \mathbf{p})$ which leads to the modified posterior. To present this prior, we first define the likelihood with a fixed $\mathbf{G} \in \boldsymbol{\mathcal{G}}$ by:
\begin{equation*}
\mathcal{L}_{\boldsymbol{G}}(\boldsymbol{\mu}, \boldsymbol{\sigma}, \mathbf{p}) \defeq  f(\mathbf{x}|\boldsymbol{\mu}, \boldsymbol{\sigma}, \mathbf{G}) f(\mathbf{G}|\mathbf{p}) = \prod_{i=1}^K \prod_{j:G_j = i}p_i f_{\mathcal{N}}(x_j \semisp \mu_i, \sigma^2_i)
\end{equation*}
Then the \cite{wasserman_asymptotic_2000} modified version of $\pi(\boldsymbol{\mu}, \boldsymbol{\sigma}, \mathbf{p})$ is:
\begin{equation}\label{prior_wasserman}
\pi^{\star}(\boldsymbol{\mu}, \boldsymbol{\sigma}, \mathbf{p}) \propto \frac{\sum_{\mathbf{G} \in \boldsymbol{\mathcal{G}}^\star} \mathcal{L}_{\mathbf{G}}(\boldsymbol{\mu}, \boldsymbol{\sigma}, \mathbf{p})}{\sum_{\mathbf{G} \in \boldsymbol{\mathcal{G}}}\mathcal{L}_{\mathbf{G}}(\boldsymbol{\mu}, \boldsymbol{\sigma}, \mathbf{p})} \pi_J(\boldsymbol{\mu}, \boldsymbol{\sigma})\pi_D(\mathbf{p}) 
\end{equation}
(compare with equation 17 in \citep{wasserman_asymptotic_2000}).
If we use the representation \eqref{post_G_expansion}, we can readily see that this indeed gives the modified posterior \eqref{good_posterior}.
%
However, the prior \eqref{prior_wasserman} is formally incorrect in a Bayesian framework, as the likelihood terms depend on the data $\mathbf{x}$. To avoid this data-dependence, we can reparametrize the model. First of all, note that the marginal posterior $f^{\star}(\boldsymbol{\mu}, \boldsymbol{\sigma} | \mathbf{x})$ from \eqref{good_posterior} is given by:
\begin{align*}
f^{\star}(\boldsymbol{\mu}, \boldsymbol{\sigma} | \mathbf{x}) \propto & \sum_{\mathbf{G} \in \boldsymbol{\mathcal{G}}^{\star}}  f(\mathbf{x} | \boldsymbol{\mu}, \boldsymbol{\sigma}, \mathbf{G}) \int f(\mathbf{G}|\mathbf{p}) \pi_D(\mathbf{p}) \mathrm{d}\mathbf{p} \; \pi_J(\boldsymbol{\mu}, \boldsymbol{\sigma})\\
\propto & \sum_{\mathbf{G} \in \boldsymbol{\mathcal{G}}}  f(\mathbf{x} | \boldsymbol{\mu}, \boldsymbol{\sigma}, \mathbf{G}) \ind_{\{\mathbf{G} \in \boldsymbol{\mathcal{G}}^\star \}}\left(\prod_{i=1}^K \Gamma (n_i(\mathbf{G}) + \delta_i) \right) \pi_J(\boldsymbol{\mu}, \boldsymbol{\sigma})
\end{align*}
where we use \eqref{standard_z_prior}. We reparametrize the mixture model to obtain this marginal posterior by placing the following prior directly on $\mathbf{G}$:
\begin{equation}\label{new_G_prior}
\pi^\star(\mathbf{G}) \propto \ind_{\{\mathbf{G} \in \boldsymbol{\mathcal{G}}^\star \}}\prod_{i=1}^K \Gamma (n_i(\mathbf{G}) + \delta_i) 
\end{equation}
Then the full posterior is given by:
\begin{equation}
f^{\star}(\boldsymbol{\mu}, \boldsymbol{\sigma}, \mathbf{G} | \mathbf{x}) \propto f(\mathbf{x} | \boldsymbol{\mu}, \boldsymbol{\sigma}, \mathbf{G}) \pi^\star(\mathbf{G})  \pi_J(\boldsymbol{\mu}, \boldsymbol{\sigma})
\end{equation}
By using the prior \eqref{new_G_prior}, we leave out the hyperparameter $\mathbf{p}$ entirely. Therefore, the resulting model is no longer equivalent to the original mixture model \eqref{mixture_1D}. However, we can show that at least marginally (for a single data point x), this model closely resembles \eqref{mixture_1D}. First, we define $\mathbf{n}(\mathbf{G}) = (n_1(\mathbf{G}),\dots n_K(\mathbf{G}))$. We condition on $\mathbf{n}(\mathbf{G}) = \mathbf{n}$, on $\boldsymbol{\mu}$, and on $\boldsymbol{\sigma}$.  We assume that $N$ data points are generated but consider only one of these, $x_1 \eqdef x$, marginalizing the others out to obtain:
\begin{equation}\label{mixture_1D_new}
f^{\star}(x | \boldsymbol{\mu}, \boldsymbol{\sigma}, \mathbf{n}) = \sum_{i=1}^K \frac{n_i}{N} f_{\mathcal{N}}(x\; ; \mu_i, \sigma^2_i)\\
\end{equation}
with the $n_i$ fixed, and satisfying:
\begin{align*}
& n_i \geq  2 \forallsp i \in 1 \dots K\\
& \sum_{i=1}^K n_i = N
\end{align*}
In this marginal density, instead of $\mathbf{p}$ we have the parameter $\mathbf{n}(\mathbf{G})/N$. $\mathbf{n}(\mathbf{G})/N$  gives the proportions with which the different components contribute to the data, whereas $\mathbf{p}$ gave the probabilities that a given data point is generated by each component. In addition, the new model directly requires each component to have at least 2 associated data points. This simply means that we require each component in the model to have actually made a meaningful contribution to the observed data. This seems to be a sensible prior assumption. In a sense, it makes the mixture fitting problem less ill-posed \citep{marin_bayesian_2005}: in the original mixture model, we could always fit models with an arbitrarily large number of components, and correspondingly low component probabilities $p_i$.\\
\indent In this new model, the $x_j,\; j \in \{1, \dots N\}$ are not independent of each other when we condition on $\mathbf{n}(\mathbf{G})$, $\boldsymbol{\mu}$ and $\boldsymbol{\sigma}^2$ and marginalize out $\mathbf{G}$. By contrast, in the standard model the $x_j$ are independent when we condition on $\mathbf{p}$, $\boldsymbol{\mu}$ and $\boldsymbol{\sigma}^2$.  The reason for this dependence is that the distribution \eqref{new_G_prior} imposes dependence between the $G_j$ (when we condition on $\mathbf{n}(\mathbf{G})$).\\
\indent As before, the prior on $\mathbf{G}$ is parameterized by $\boldsymbol{\delta} = (\delta_1,\dots \delta_K)$. If we make the choice $\boldsymbol{\delta} = (1, \dots 1)$, then $\pi^{\star}(\mathbf{G})$ is uniform in terms of the numbers of points assigned to the different components, when we consider only assignments with at least $2$ points per component. That is, all $\mathbf{n}(\mathbf{G})$ with every component $\geq 2$ have the same prior probability. This resembles the behavior of the standard prior \eqref{standard_z_prior} when $\boldsymbol{\delta} = (1, \dots 1)$.

\subsection{Implementations}\label{Section_imp}

We developed Markov Chain Monte Carlo methods for sampling from the posterior of the GMM with the priors $\pi^{\star}(\mathbf{G})$ \eqref{new_G_prior} and $\pi_J(\boldsymbol{\mu}, \boldsymbol{\sigma})$ \eqref{Jeffreys_prior}. For better comparison with the computational results in section \ref{section_weakinf}, we at first continued to use a Gibbs sampling-based method. Note that use of the prior $\pi^{\star}(\mathbf{G})$ does not mean that we have to perform Gibbs sampling on a higher-dimensional space, as Gibbs sampling techniques for the standard model \eqref{mixture_1D} also require $\mathbf{G}$ to be explicitly sampled. Nevertheless, optimizing the efficiency of the Gibbs sampler still seemed worthwhile. To do so, we implemented a scheme which only samples from $\mathbf{G}$, by Gibbs sampling from the posterior $f(\mathbf{G}|\mathbf{x})$ with $\boldsymbol{\mu}$ and $\boldsymbol{\sigma}^2$ integrated out. This is an example of collapsed Gibbs sampling, which in theory should converge faster than basic Gibbs sampling \citep{liu_collapsed_1994}.\\
\indent $f(\mathbf{G}|\mathbf{x})$ is given by \eqref{z_post}, with the prior $\pi(\mathbf{G})$ \eqref{standard_z_prior} replaced by $\pi^\star(\mathbf{G})$ \eqref{new_G_prior}, and $\pi(\mu_i, \sigma_i) \propto \sigma_i^{-1}$ in accordance with the Jeffreys prior ($\sigma_i$, not $\sigma_i^2$, is now the integration variable). The integrals over $\mu_i$ and $\sigma_i$ which appear in the expression for $f(\mathbf{G}|\mathbf{x})$ are given by:
\begin{align*}
& \int \int \left( \prod_{j:G_j=i}  f_\mathcal{N}(x_j \semisp \mu_i, \sigma^2_i)\right) \sigma_i^{-1}\mathrm{d} \mu_i \mathrm{d} \sigma_i\\
& \propto\;  \pi^{\frac{1-n_i}{2}} \left( \frac{1}{n_i}\sum_{j:G_j = i} x_j^2 - \frac{1}{n_i^2}\left(\sum_{j:G_j=i} x_j\right)^2\right)^{\frac{1-n_i}{2}}n_i^{-\frac{n_i}{2}} \Gamma\left( \frac{n_i - 1}{2} \right)
\end{align*}\\
where $n_i = n_i(\mathbf{G})$ (see appendix, \ref{int_1D}). The collapsed Gibbs sampling algorithm then is as follows: to generate $S$ samples $\mathbf{G}^{(s)},\; s \in \{1, \dots S\}$, from a Markov chain whose distribution converges to $f(\mathbf{G}|\mathbf{x})$, we run:\\
\begin{align*}
&\mathrm{Initialize}\; \mathbf{G}^{(0)} \mathrm{\;at\;random\;such\;that\;} \pi^\star(\mathbf{G}^{(0)}) > 0\\
&\mathrm{For}\; s \in \{1, \dots S\}\\
&\quad \mathrm{For} \; j \in \{1, \dots N\}\\
&\quad\quad \mathrm{Sample}\; G_j^{(s)} \;\mathrm{from}\; f(G_j|G_1^{(s)}, \dots G_{j-1}^{(s)}, G_{j+1}^{(s-1)}, \dots G_{N}^{(s-1)}, \mathbf{x})\\
&\quad \mathrm{End}\\
&\mathrm{End}\\
\end{align*}
The density $f(G_j|G_1^{(s)}, \dots G_{j-1}^{(s)}, G_{j+1}^{(s-1)}, \dots G_{N}^{(s-1)}, \mathbf{x})$ is obtained readily as it is proportional to $f(\mathbf{G}|\mathbf{x})$. We can discard a burn-in period, then take e.g.\ every 10th sample to obtain an approximate sample from $f(\mathbf{G}|\mathbf{x})$. For each $\mathbf{G}$ within this sample, we can obtain a sample for each pair $(\mu_i, \sigma_i^2),\; i\in \{1, \dots K\}$ from their joint posterior conditioned on $\mathbf{G}$ and $\mathbf{x}$ (this density is simply normal-inverse gamma). The final result is an approximate sample from $f(\boldsymbol{\mu}, \boldsymbol{\sigma}^2|\mathbf{x})$. Note that we obtain these samples retrospectively, they are not part of the actual Markov chain. We can of course easily obtain samples from the component proportions $n_i/N$, given the set of samples of $\mathbf{G}$.\\
\indent This scheme can also be used with a normal-inverse gamma prior on each pair $(\mu_i, \sigma^2_i), \; i \in \{1, \dots K\}$. In this case, the integrals in \eqref{z_post} are given by $f_i(\mathbf{x}, \mathbf{G})$ as in \eqref{fi_expression}. Besides the modified prior $\pi^\star(\mathbf{G})$ \eqref{new_G_prior}, we can also use the original prior $\pi(\mathbf{G})$  \eqref{standard_z_prior}. In the latter case, for each sample $\mathbf{G}$, we can obtain a sample of $\mathbf{p}$ from its distribution conditioned on $\mathbf{G}$ (which is Dirichlet). Because the $\mu_i$ and $\sigma^2_i$ must be integrated out, there are some limitations on the priors that can be used for these parameters (the integral $\int \int f(\mu_i, \sigma_i | \mathbf{x}, \mathbf{G}) \mathrm{d}\mu_i \mathrm{d} \sigma_i$ must be available in closed form). In particular, the \citet{richardson_bayesian_1997} hierarchical model cannot be implemented.\\
\indent We found that although the Gibbs sampling-based method performed adequately for some example datasets, on others it failed to converge in a reasonable number of iterations, as demonstrated by the fact that the approximation of the posterior of some $\mu_i$ did not show the expected symmetric modes. This is a well-known issue with Gibbs sampling for mixture models \citep{jasra_markov_2005, marin_bayesian_2005}. An alternative is to use the Metropolis-Hastings algorithm, either in a standard form \citep{marin_bayesian_2005} or in a tempering MCMC scheme \citep{jasra_markov_2005}. This tends to explore the posterior density better, as demonstrated by the fact that switching between posterior modes occurs more frequently. Therefore, we implemented a simple Metropolis-Hastings-based scheme for our model. This method had faster convergence on test data, and is probably more suitable for most practical applications. Details of the implementation are in the appendix \eqref{MH_imp}. MATLAB and R code for both collapsed Gibbs and Metropolis-Hastings sampling is provided at \texttt{https://sourceforge.net/projects/bayesiangmm/}.

\subsection{Comparison of the models}\label{Section_comp}

\begin{figure}
\vspace*{0cm}\hspace*{-2cm}\includegraphics[scale=0.75,clip, trim = 0mm 102mm 0mm 90mm]{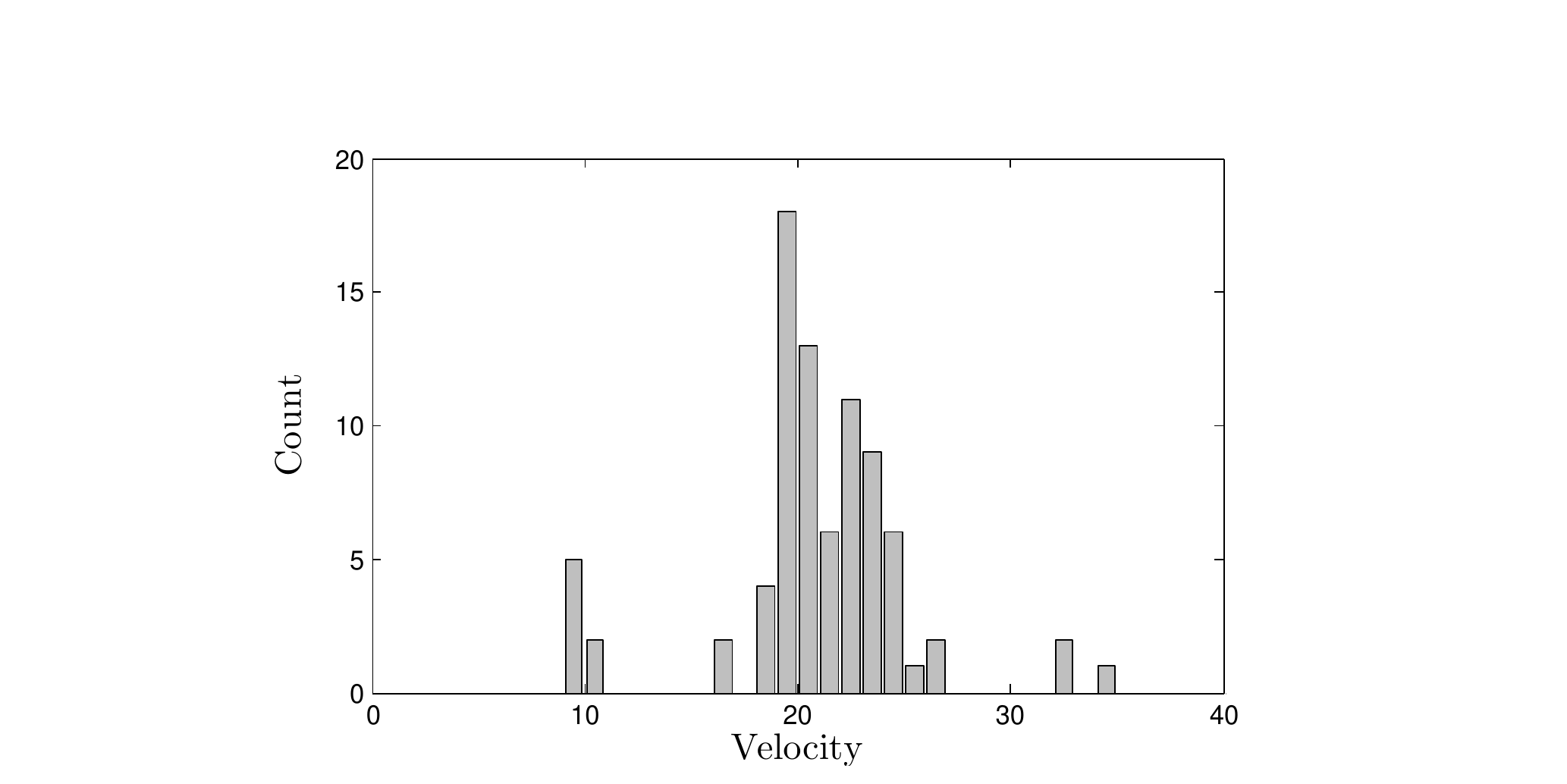}
\caption{Histogram of the galaxy dataset.}\label{galaxy}
\end{figure}

\begin{figure}
\vspace*{0cm}\hspace*{-2cm}\includegraphics[scale=0.75,clip, trim = 0mm 100mm 0mm 90mm]{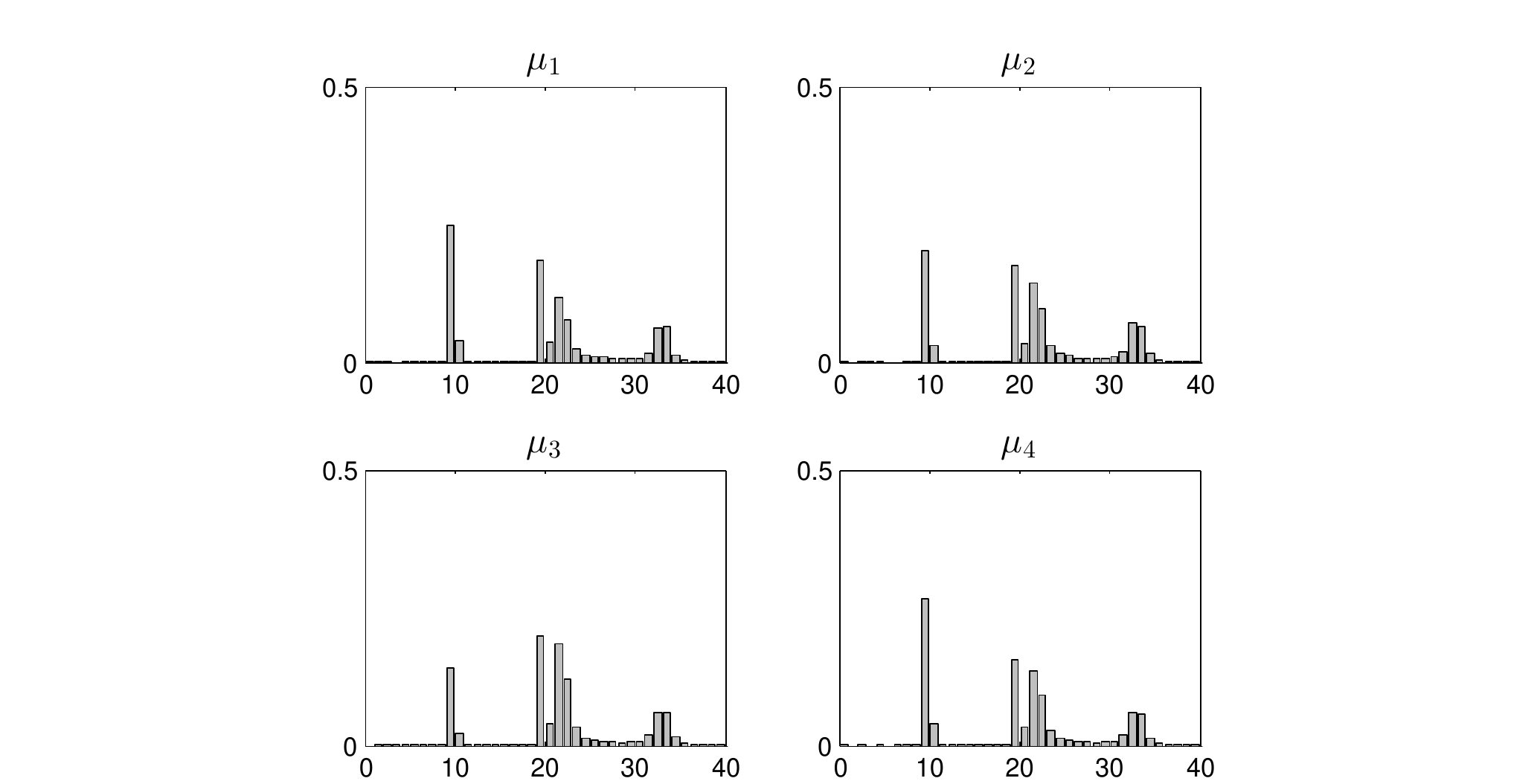}
\caption{Samples from the posteriors of $\mu_1$ and $\mu_2$, from fitting the standard GMM \eqref{mixture_1D} with $K=4$ to the galaxy dataset. The normal-inverse gamma prior was used, with $\alpha=0.01,\; \beta=0.01,$ and $\kappa=0.01$. The burn-in period was $10^4$ steps, post-burn-in $5 \times 10^5$, every 10-th state was saved.}\label{galaxy_analysis_nivg}
\vspace*{0cm}\hspace*{-2cm}\includegraphics[scale=0.75,clip, trim = 0mm 100mm 0mm 90mm]{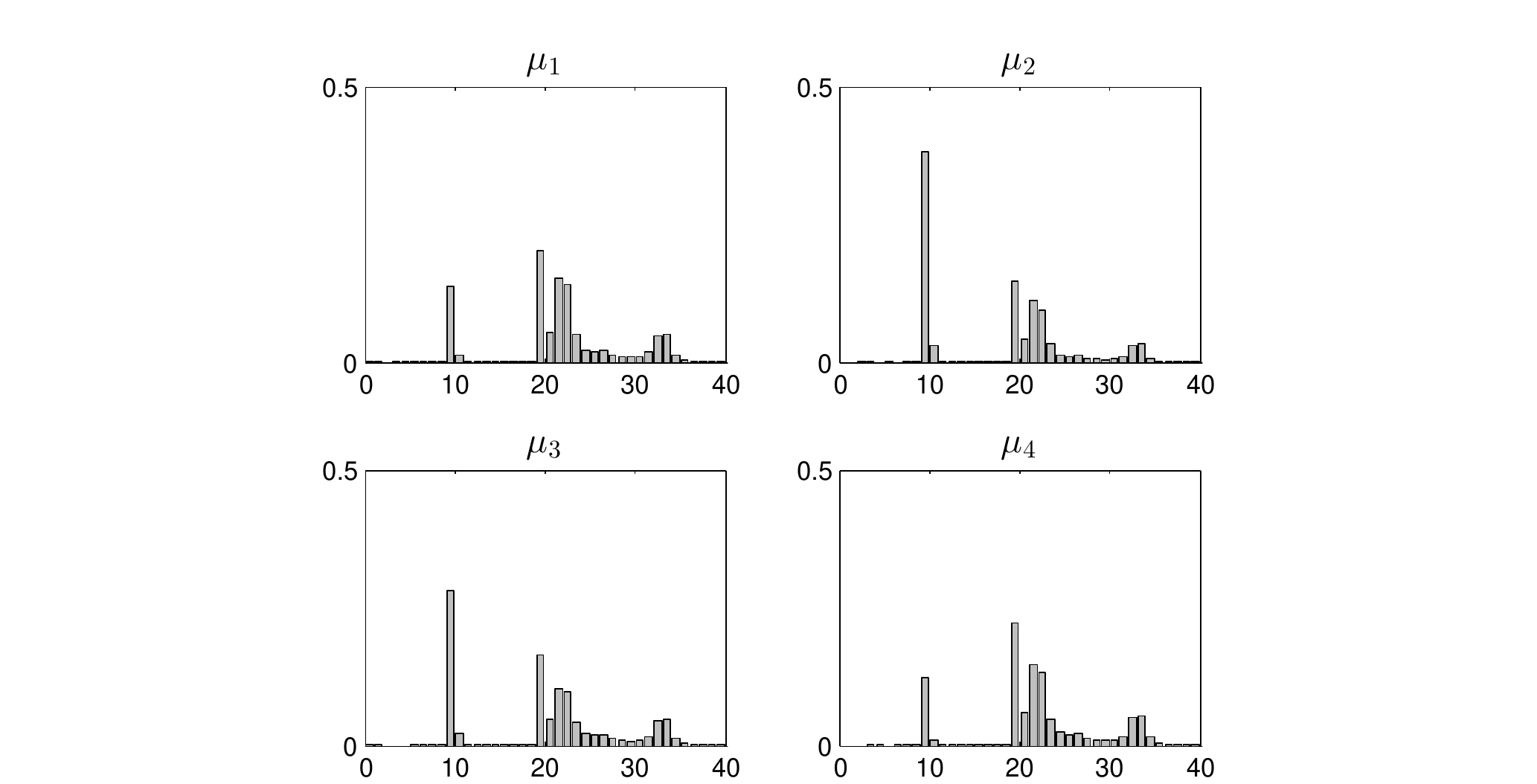}
\caption{Samples from the posteriors of $\mu_1$ and $\mu_2$, from fitting a GMM with $K=4$ to the galaxy dataset, using the priors $\pi^\star(\mathbf{G})$ \eqref{new_G_prior} and $\pi_J(\boldsymbol{\mu}, \boldsymbol{\sigma})$ \eqref{Jeffreys_prior}. The burn-in was $10^4$ steps, post-burn-in $5 \times 10^5$, every 10-th state was saved.}\label{galaxy_analysis_improper}
\end{figure}

\begin{figure}
\vspace*{-4cm}\hspace*{-2cm}\includegraphics[scale=0.75,clip, trim = 0mm 120mm 0mm 110mm]{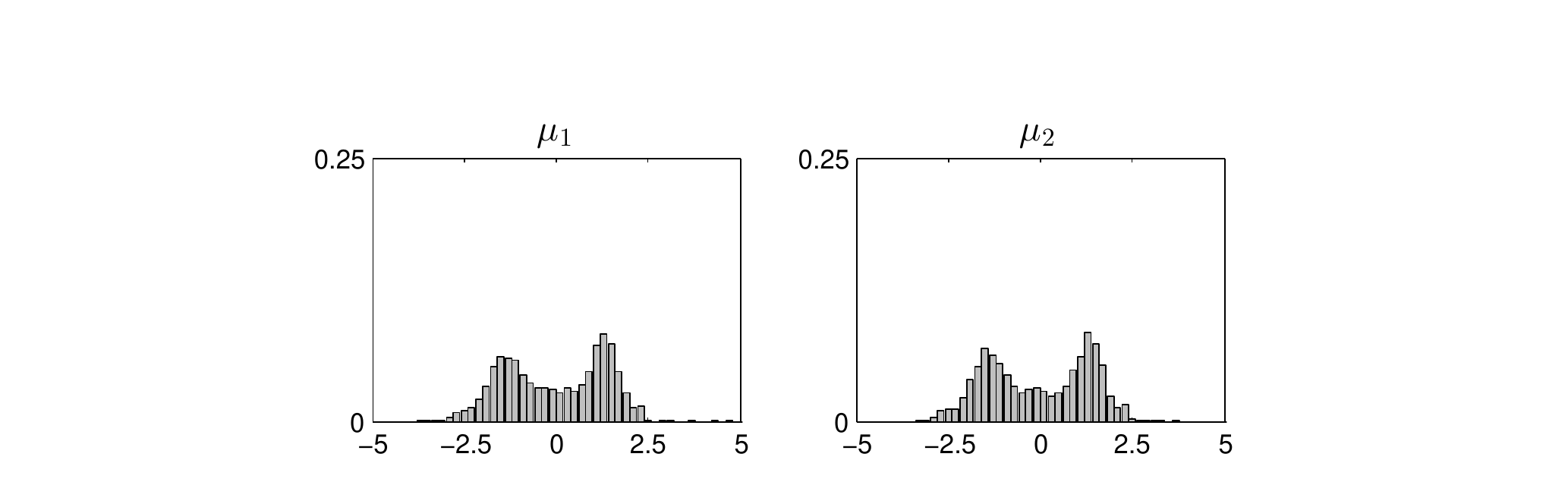}
\caption{Samples from the posteriors of $\mu_1$ and $\mu_2$ using a GMM with $K=2$, for the synthetic data from figure \ref{synth_data}. The priors $\pi^\star(\mathbf{G})$ \eqref{new_G_prior} and $\pi_J(\boldsymbol{\mu}, \boldsymbol{\sigma})$ \eqref{Jeffreys_prior} were used. The burn-in was $10^4$ steps, post-burn-in $10^5$, every 10-th state of the chain was saved. Compare with figures \ref{synth_data_nivg} and \ref{synth_data_rg} which show the same inference with the standard GMM given by \eqref{mixture_1D}, using weakly informative proper priors.}\label{synth_data_imp}
\vspace*{0cm}\hspace*{-2cm}\includegraphics[scale=0.75,clip, trim = 0mm 110mm 0mm 100mm]{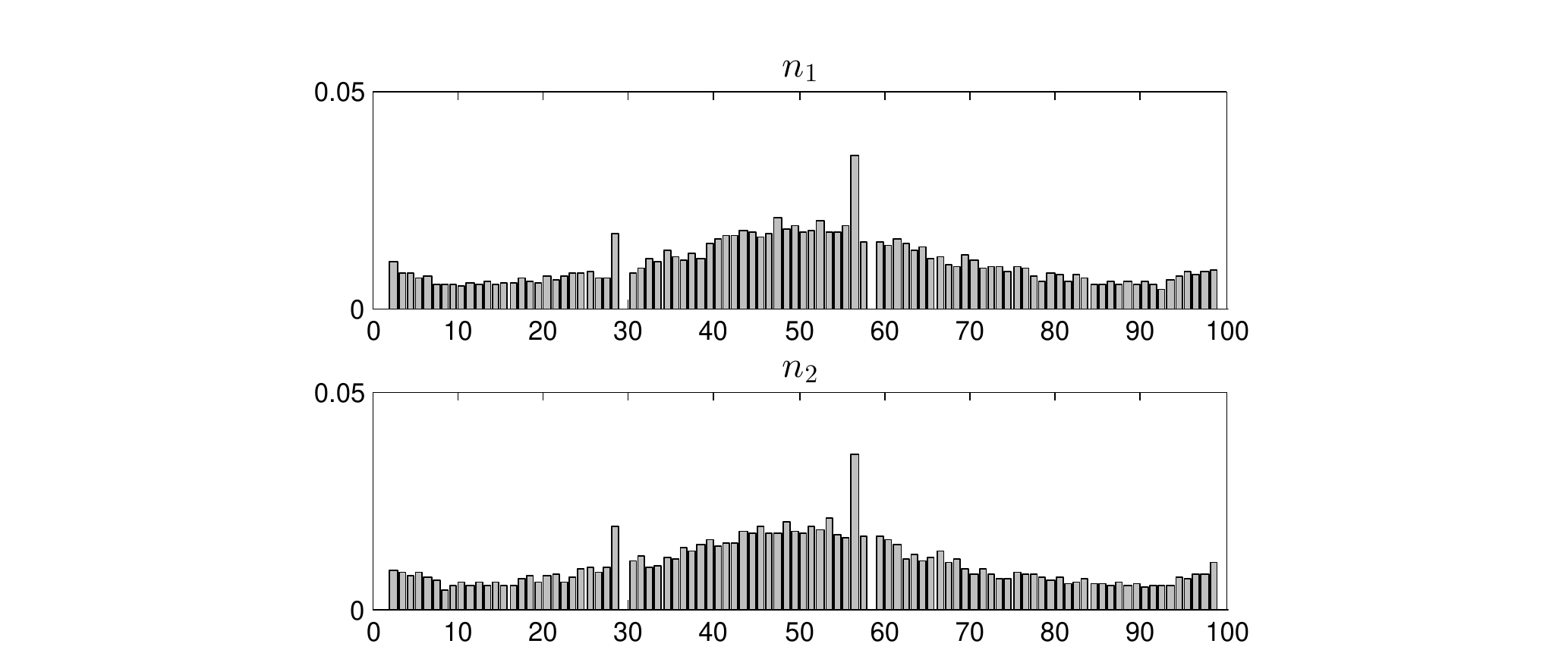}
\caption{Samples from the posteriors of $n_1$ and $n_2$ from the inference shown in figure \ref{synth_data_imp}. }\label{synth_data_imp_groups}
\end{figure}

\indent We used a collapsed Gibbs sampling scheme to implement the GMM with the priors $\pi^\star(\mathbf{G})$ \eqref{new_G_prior} and $\pi_J(\boldsymbol{\mu}, \boldsymbol{\sigma})$ \eqref{Jeffreys_prior}. For a comparison, we used the same method to implement the standard GMM \eqref{mixture_1D} with a Dirichlet prior on $\mathbf{p}$ and normal-inverse gamma priors on the pairs $(\mu_i, \sigma^2_i)$, $i \in \{1, \dots K\}$. We first tested these models on the galaxy dataset, which is a widely used dataset originally analyzed by \cite{roeder_density_1990}. The dataset consists of 82 points (the velocities of different galaxies), a histogram of which is shown in figure \ref{galaxy}. We fit the data with a mixture of 4 Gaussians, using both models. Note that 4 was chosen mainly as an example, we do not assume that it is the most appropriate number. The resulting posterior densities are very similar (figures \ref{galaxy_analysis_nivg}, \ref{galaxy_analysis_improper}). Therefore, on this data set with mostly well-separated modes, our model gives the same result as the standard model. Note that for each model, all 4 component means have similar posterior densities, indicating that the Gibbs sampler was able to move between different modes of the posterior likelihood.\\
\indent We next used the new model to analyze the synthetic dataset from figure \ref{synth_data}. The resulting posteriors for $\mu_1$ and $\mu_2$ are shown in figure \ref{synth_data_imp}. They have only two distinct modes, similar to the result from the standard model with relatively large values for the prior parameters (c.f. figures \ref{synth_data_nivg} and \ref{synth_data_rg}). For this analysis, we also examined the posterior distribution of $n_1$ and $n_2$. We found that there does not seem to be a significant peak for $n_i$ near $2$. This suggests that constraining $n_i$ to be at least $2$ did not significantly affect the posterior, other than removing the enrichment of smaller $n_i$ values seen for the standard model (c.f. figures \ref{synth_data_nivg_zeros} and \ref{synth_data_rg_zeros}).

\section{Conclusion}

We have shown that improper priors can be used for Bayesian inference of GMMs with only a slight modification of the original model, which consists of using the modified prior $\pi^\star(\mathbf{G})$ \eqref{new_G_prior}, and recasting inference in terms of proportions $\mathbf{n}/N$ rather than probabilities $\mathbf{p}$. Our approach is generic: any improper prior on the parameters of a Gaussian distribution can be used, as long as a specific number of data points is guaranteed to make the posterior $P - a.s.$ proper. Besides the Jeffreys prior mentioned thus far, uniform priors for variance parameters would also be possible \citep{gelman_prior_2006}. Also, our approach can be generalized to more than one dimension, although the minimum number of data points per component will generally need to be increased from 2. This modification of the model has several advantages. First of all, we can use noninformative priors to avoid the problem of parameter choice for weakly informative proper priors. This is crucial, as we have seen that proper priors are prone to a prior domination effect which makes parameter choice difficult. Using a model which permits both proper and improper priors also enables the effect of informative proper priors on the posterior to be compared with a noninformative improper prior.\\
\indent Throughout this study, we have held the number of components ($K$) fixed. Methods which treat this number as random are very useful for nonparametric density estimation - these have been studied by \citep{ferguson_bayesian_1983, escobar_bayesian_1995, richardson_bayesian_1997, stephens_bayesian_2000}, among others. However, suppose we not only want to obtain a predictive density, but also want to infer the parameters of the different mixture components. Then treating $K$ as random leads to some complications.  For one, proper priors on the component parameters must be used in this case, but estimates of $K$ tend to be rather sensitive to the choice of these priors. This is true both for methods using reversible jump MCMC \citep{richardson_bayesian_1997} and birth-death process-based methods \citep{stephens_bayesian_2000}. Also, a high degree of uncertainty about the number of components may remain \citep{escobar_bayesian_1995, richardson_bayesian_1997, stephens_bayesian_2000}. This can be a problem if our strategy is to infer the component parameters conditional on the maximum a posteriori number of components. Thus, it may be more useful to take a range of possible values of $K$ (maybe based on prior knowledge), and sample from the posterior of a mixture model for each of these values. At the very least, this has the advantage of preserving as much information as possible.\\

\section{Acknowledgements}

I would like to thank Peter B{\"u}hlmann and Hans Rudolf K{\"u}nsch for their very helpful comments and advice.

\section{Appendix}

\subsection{Proof of Lemma 1}\label{lemma_1}

\noindent \textbf{Lemma 1}\\

\noindent Assume that $N > K$, and that $\alpha = c_1 \kappa$ and $\beta = c_2 \kappa$, with fixed constants $c_1 > 0$ and $c_2 > 0$. Let $\mathbf{G'}, \mathbf{G''} \in \boldsymbol{\mathcal{G}}$ be vectors of latent variables. If $\mathbf{G'}$ assigns all data points to a single component, i.e. $\#\{i \in \{1, \dots K\} : n_i(\mathbf{G'}) = 0\} = K - 1$, and $\mathbf{G''}$ does not do this, then $P-a.s.$:

\begin{displaymath}
\lim_{\kappa \to 0} \frac{f(\mathbf{G'}|\mathbf{x})}{f(\mathbf{G''}|\mathbf{x})} = \infty
\end{displaymath}\\

\noindent\textit{Proof:} \\
\\
We have (see \eqref{fi_notation}):

\begin{equation}\label{G_post_exp}
f(\mathbf{G}|\mathbf{x}) \propto \pi(\mathbf{G}) \prod_{i=1}^K f_i(\mathbf{x} , \mathbf{G}) 
\end{equation}
with:
\begin{align*}
f_i(\mathbf{x},\mathbf{G})= \; \frac{(2\beta)^\alpha \kappa^{\frac{1}{2}}\;\Gamma\left(\frac{n_i}{2} + \alpha\right)}{\pi^{\frac{n_i}{2}} (n_i + \kappa)^{\frac{n_i + 1}{2} + \alpha}\;\Gamma(\alpha)} \left(\frac{1}{n_i + \kappa}\sum_{j:G_j=i} x_j^2 -\left(\frac{1}{n_i + \kappa}\sum_{j:G_j=i} x_j\right)^2 + \frac{2\beta}{n_i + \kappa}\right)^{-\frac{n_i}{2} -\alpha}\\
\end{align*}
where $n_i = n_i(\mathbf{G})$. For $n_i(\mathbf{G}) = 0$, we have $f_i(\mathbf{x},\mathbf{G}) = 1$.
For $n_i(\mathbf{G}) = 1$, we have:
\begin{displaymath}
f_i(\mathbf{x},\mathbf{G}) = \frac{(2c_2)^{c_1 \kappa} \Gamma\left( \frac{1}{2} + c_1\kappa \right)}{\pi^{\frac{1}{2}} (1 + \kappa)^{\frac{1}{2}}\Gamma(c_1 \kappa)} \left( \frac{1}{1 + \kappa}\sum_{j: G_j = i} x_j^2 + 2c_2  \right)^{-\frac{1}{2} - c_1\kappa}
\end{displaymath}
Therefore, for $n_i(\mathbf{G}) = 1$, $f_i(\mathbf{x} , \mathbf{G}) = \mathcal{O}\left(\kappa\right)$ and $f_i(\mathbf{x} , \mathbf{G})^{-1} = \mathcal{O}\left(\kappa^{-1}\right)$ as $\kappa \to 0$.
For $n_i(\mathbf{G}) > 1$, we have:
\begin{displaymath}
f_i(\mathbf{x} , \mathbf{G}) = \frac{(2c_2)^{c_1 \kappa} \kappa^{\frac{1}{2} + c_1\kappa}\Gamma\left( \frac{n_i}{2} +  c_1\kappa \right)}{\pi^{\frac{n_i}{2}} (n_i + \kappa)^{\frac{1}{2}}\Gamma(c_1 \kappa)} \left[ (n_i + \kappa)V_i(\mathbf{x} , \mathbf{G},  \kappa) + 2c_1\kappa \right]^{-\frac{n_i}{2} - c_1\kappa}
\end{displaymath}
with:
\begin{equation}\label{V_def}
V_i(\mathbf{x}, \mathbf{G},  \kappa) \defeq \frac{1}{n_i + \kappa}\sum_{j:G_j = i} x_j^2 -\left(\frac{1}{n_i + \kappa}\sum_{j:G_j=i} x_j\right)^2
\end{equation}
$V_i(\mathbf{x} , \mathbf{G},  \kappa) \geq 0$ by the Cauchy-Schwarz inequality.
We now use the assumption that $\mathbf{x} \in \mathbb{R}^N$ consists of i.i.d. samples from a distribution $F$ that is dominated by Lebesgue measure on $\mathbb{R}$. For any $S \subseteq \{1, \dots N\}$, let $\mathbf{x}(S) \in \mathbb{R}^{\#S}$ denote a vector consisting of all $x_j$ with $j \in S$. We define: 
\begin{equation*}
V(\mathbf{y}) = \frac{1}{n}\sum_{j=1}^{n} y_j^2 -\left(\frac{1}{n}\sum_{j=1}^{n} y_j\right)^2
\end{equation*}
for $\mathbf{y} \in \mathbb{R}^n$.
Now, we have:
\begin{equation*}
\mathrm{P}\left( \exists S \subseteq \{1, \dots N\}: \#S > 1, V\left( \mathbf{x}(S) \right) = 0 \right) \leq \sum_{S \subseteq \{1, \dots N\}, \#S > 1} \mathrm{P}\left(V\left( \mathbf{x}(S) \right) = 0\right)\\
\end{equation*}
For $n > 1$, we have by Tonelli's theorem:
\begin{equation*}
\int_{\mathbb{R}^n} \ind_{\{V(\mathbf{y}) = 0\}}\mathrm{d}\mathbf{y} = 0
\end{equation*}
Therefore, since $F$ is dominated by Lebesgue measure on $\mathbb{R}$, $\mathrm{P}\left(V\left( \mathbf{x}(S) \right) = 0\right) = 0$ for all $S \subseteq \{1, \dots N\}$ with $\#S > 1$. This means that $P-a.s.$, $\forall\; i \in \{1, \dots K\}$ and $\forall\; \mathbf{G} \in \boldsymbol{\mathcal{G}}$ with $n_i(\mathbf{G}) > 1$, we have $V_i(\mathbf{x} , \mathbf{G}, 0) > 0$.
Now we obtain for $n_i(\mathbf{G}) > 1$, $P-a.s.$: $f_i(\mathbf{x} , \mathbf{G}) = \mathcal{O}\left(\kappa^{\frac{3}{2}}\right)$ and $f_i(\mathbf{x} , \mathbf{G})^{-1} = \mathcal{O}\left(\kappa^{-\frac{3}{2}-\epsilon}\right)$ for arbitrarily small $\epsilon > 0$, as $\kappa \to 0$.
Using \eqref{G_post_exp}, we have:
\begin{displaymath}
\frac{f(\mathbf{G''}|\mathbf{x})}{f(\mathbf{G'}|\mathbf{x})} = \frac{\pi(\mathbf{G''})}{\pi(\mathbf{G'} )} \frac{\prod_{i=1}^K f_i(\mathbf{x} , \mathbf{G''})}{\prod_{i=1}^K f_i(\mathbf{x} , \mathbf{G'})}
\end{displaymath}
$\prod_{i=1}^K f_i(\mathbf{x} , \mathbf{G'})^{-1}$ is $P-a.s.$ $\mathcal{O}\left(\kappa^{-\frac{3}{2}-\epsilon}\right)$ as $\kappa \to 0$ (because $\mathbf{G}'$ assigns all data points to one component). $\prod_{i=1}^K f_i(\mathbf{x} , \mathbf{G''})$ is $P-a.s.$ at most $\mathcal{O}\left(\kappa^{\frac{3}{2} + 1}\right)$ as $\kappa \to 0$. This is the case when it assigns all but one data point to one component, and one data point to another component, and none to the other components (note that this depends on the assumption $N > K$). Therefore $\frac{f(\mathbf{G''}|\mathbf{x})}{f(\mathbf{G'}|\mathbf{x})}$ is $P-a.s.$ $\mathcal{O}\left(\kappa^{1-\epsilon}\right)$ as $\kappa \to 0$, so its limit is $P-a.s.$ zero, and we obtain the statement of the lemma.\\
\indent In the latter part of the simulations (see figure \ref{synth_data_nivg}), we held $\alpha$ fixed at a small value to avoid computational difficulties. Using the same approach as above, we can show that in the case of two components, with fixed $\alpha$ and $\beta = c_1\kappa$, if $\mathbf{G''}$ assigns each component more than one data point and $\mathbf{G'}$ does not do this, then $P-a.s.$ $\frac{f(\mathbf{G'}|\mathbf{x})}{f(\mathbf{G''}|\mathbf{x})} \to \infty$ as $\kappa \to 0$.

\subsection{Integrating out the Jeffreys prior}\label{int_1D}

Here, we derive the closed-form expression for 
\begin{equation*}
f_i(\mathbf{x}, \mathbf{G}) \propto \int \int \left( \prod_{j:G_j=i} f_\mathcal{N}(x_j \semisp \mu_i, \sigma_i)\right) \pi(\mu_i, \sigma_i)\mathrm{d} \mu_i \mathrm{d} \sigma_i
\end{equation*}
with an improper Jeffreys prior $\pi(\mu_i, \sigma_i)\propto \sigma_i^{-1}$. We assume that $n_i(\mathbf{G}) > 1$ holds, and that $\mathbf{x}$ was generated via independently drawing each $x_j, j \in \{1, \dots N\}$ from some distribution on $\mathbb{R}$ that is absolutely continuous with respect to Lebesgue measure. Then, we can readily derive the following:
\begin{align*}
f_i(\mathbf{x}, \mathbf{G}) \defeq & \int \int (2\pi)^{-\frac{n_i}{2}} \sigma_i^{-n_i-1} \exp \left( -\frac{1}{2 \sigma_i^2}\sum_{j:G_j=i} \left(\mu_i - x_j \right)^2 \right)\mathrm{d} \mu_i \mathrm{d} \sigma_i\\
%
 %
 %
\propto & \int (2\pi)^{-\frac{n_i - 1}{2}} \sigma_i^{-n_i} {n_i}^{-\frac{1}{2}} \exp\left(-\frac{n_i}{2\sigma_i^2}\left(\frac{1}{n_i}\sum_{j:G_j=i} x_j^2 - \left(\frac{1}{n_i}\sum_{j:G_j=i} x_j\right)^2 \right)\right)\mathrm{d} \sigma_i\\
\end{align*}
Define
\begin{equation*}
V_i(\mathbf{x}, \mathbf{G}) \defeq \frac{1}{n_i}\sum_{j:G_j=i} x_j^2 - \left(\frac{1}{n_i}\sum_{j:G_j=i} x_j\right)^2
\end{equation*}
%
%
%
$V_i(\mathbf{x}, \mathbf{G})$ is $P - a.s. > 0$ for $n_i(\mathbf{G}) > 2$, using the assumption on the distribution of $\mathbf{x}$ (see the proof of Lemma 1). Assuming $V_i(\mathbf{x}, \mathbf{G}) > 0$ holds, we make the substitution $u = \frac{n_i V_i(\mathbf{x}, \mathbf{G})}{2\sigma_i^2}$, and solve the integral to obtain:
%
%
%
\begin{equation}\label{x_marginal}
f_i(\mathbf{x}, \mathbf{G}) \propto\;  (\pi V_i(\mathbf{x}, \mathbf{G}))^{\frac{1-n_i}{2}}{n_i}^{-\frac{n_i}{2}} \; \Gamma\left( \frac{n_i - 1}{2} \right)
\end{equation}

\subsection{Metropolis-Hastings implementation of our model}\label{MH_imp}

For increased computational efficiency, we implemented the GMM with the priors $\pi^\star(\mathbf{G})$ \eqref{new_G_prior} and $\pi_J(\boldsymbol{\mu}, \boldsymbol{\sigma})$ \eqref{Jeffreys_prior} using a Metropolis-Hastings algorithm. This was based on the algorithm given for the standard mixture model in \cite{marin_bayesian_2005}, but with several modifications. Our scheme produces approximate samples from the joint posterior distribution of $\boldsymbol{\mu}$, $\boldsymbol{\sigma}$ and $\mathbf{G}$, given $\mathbf{x}$. From each sample of $\mathbf{G}$, we can immediately compute $\mathbf{n}/N$, i.e. the proportions with which the different Gaussian components contribute to the observed data.\\ 
\indent The proposal distributions for $\boldsymbol{\mu}$, $\boldsymbol{\sigma}$ and $\mathbf{G}$ are all denoted by $g(\cdot | \cdot)$ in the following. We take $g(\boldsymbol{\mu}'|\boldsymbol{\mu})$ to be Gaussian with mean vector $\boldsymbol{\mu}$ and covariance proportional to the identity matrix. To make the Metropolis-Hastings random walk more efficient at exploring the posterior distribution of mean vectors, we also restrict the values of the components of $\boldsymbol{\mu}$ to a pre-specified interval $[\mu_{min}, \mu_{max}]$. Any components of the proposal $\boldsymbol{\mu}'$ which are outside of the interval are ``wrapped around'' so they are inside the interval, at its opposite end. The standard deviation of $g(\boldsymbol{\mu}'|\boldsymbol{\mu})$  may need to be adjusted for efficient exploration, depending on the range of the input data.\\ 
\indent The proposal distribution $g(\boldsymbol{\sigma}'|\boldsymbol{\sigma})$ is also Gaussian, centered at $\boldsymbol{\sigma}$ with covariance proportional to the identity matrix. The components of $\boldsymbol{\sigma}$ are restricted to be greater than a specified minimum value, $\sigma_{min}$, e.g. 0.01 (if proposal values would be smaller than this value, they are reflected around it). The standard deviation of this proposal distribution may also need to be tuned, depending on the input data. Note that because of the restrictions $\mu_i \in [\mu_{min}, \mu_{max}]$ and $\sigma_i \geq \sigma_{min}$, we do not use the true Jeffreys prior $\pi_J(\boldsymbol{\mu}, \boldsymbol{\sigma})$ \eqref{Jeffreys_prior}, but an approximation instead. The restrictions improve the mixing properties of the Markov chain. The proposal distributions for $\boldsymbol{\mu}$ and $\boldsymbol{\sigma}$ are symmetric, so they cancel in the expression for the acceptance probability $A$, in the algorithm below.\\
\indent The proposal distribution $g(\mathbf{G}'|\boldsymbol{\mu}', \boldsymbol{\sigma}')$ is taken to be proportional to the likelihood $f(\mathbf{x}|\boldsymbol{\mu}', \boldsymbol{\sigma}', \mathbf{G}')$ in the standard model \eqref{mixture_1D}:
\begin{equation*}
g(\mathbf{G'}|\boldsymbol{\mu}', \boldsymbol{\sigma}') \propto \prod_{i=1}^K \prod_{j:G'_j=i} \sigma_{i}^{'-1}\exp\left(-\frac{(x_j - \mu'_{i})^2}{2\sigma_{i}^{'2}}\right)
\end{equation*}
This is equivalent to the posterior $f(\mathbf{G} | \boldsymbol{\mu}', \boldsymbol{\sigma}', \mathbf{x})$ in the standard model, with $p_i = 1/K \forallsp i \in \{1, \dots K\}$, i.e. uniform component probabilities. Then the following algorithm simulates $S$ states of a Markov chain whose distribution converges to $f(\boldsymbol{\mu}, \boldsymbol{\sigma}, \mathbf{G} | \mathbf{x})$:\\
\begin{align*}
&\quad\mathrm{Choose\;starting\;values\;} \boldsymbol{\mu}^{(0)} \;\mathrm{and}\;  \boldsymbol{\sigma}^{(0)}\\
&\quad\mathrm{Do}:
\mathrm{Draw\;} \mathbf{G}^{(0)} \mathrm{\;from\;} g(\mathbf{G} | \boldsymbol{\mu}^{(0)}, \boldsymbol{\sigma}^{(0)})
\quad\mathrm{While\;}  \pi^\star(\mathbf{G}^{(0)}) =0 \\
\\
&\quad\mathrm{For}\; s \in \{1, \dots S\}\\
&\quad\quad \mathrm{Draw\;}   \boldsymbol{\mu}' \mathrm{\;from\;} g(\boldsymbol{\mu}' | \boldsymbol{\mu}^{(s-1)}),\;
 \boldsymbol{\sigma}' \mathrm{\;from\;} g(\boldsymbol{\sigma}' | \boldsymbol{\sigma}^{(s-1)}),\;
 \mathrm{and\;} \mathbf{G}' \mathrm{\;from\;} g(\mathbf{G}' | \boldsymbol{\mu}', \boldsymbol{\sigma}')\\
\\
&\quad\quad \mathrm{Set\;} A = \frac{g(\mathbf{G}^{(s-1)} | \boldsymbol{\mu}^{(s-1)}, \boldsymbol{\sigma}^{(s-1)}) }{g(\mathbf{G}' | \boldsymbol{\mu}', \boldsymbol{\sigma}')} \frac{f(\mathbf{x}|\boldsymbol{\mu}', \boldsymbol{\sigma}', \mathbf{G}')\pi^\star(\mathbf{G}')}{f(\mathbf{x}|\boldsymbol{\mu}^{(s-1)}, \boldsymbol{\sigma}^{(s-1)}, \mathbf{G}^{(s-1)})\pi^\star(\mathbf{G}^{(s-1)})}\\
\\
&\quad\quad \mathrm{Draw}\; u \;\mathrm{from\;the\;uniform\;distribution\;on\;} (0, 1)\\
\\
&\quad\quad \mathrm{If}\;u < A, \mathrm{then\; set:\;}
\quad\quad\quad \boldsymbol{\mu}^{(s)} = \boldsymbol{\mu}', 
\quad\quad \boldsymbol{\sigma}^{(s)} = \boldsymbol{\sigma}', 
\quad\quad \mathbf{G}^{(s)} = \mathbf{G}'\\
&\quad\quad \mathrm{Else\;set:\;}\quad
\quad\quad\quad \boldsymbol{\mu}^{(s)} = \boldsymbol{\mu}^{(s-1)}, 
\quad\quad \boldsymbol{\sigma}^{(s)} = \boldsymbol{\sigma}^{(s-1)}, 
\quad\quad \mathbf{G}^{(s)} = \mathbf{G}^{(s-1)}
\end{align*}

\bibliographystyle{authordate1}

\bibliography{bib}

\newpage

\end{document}